\pdfoutput=1
\documentclass[lettersize,journal]{IEEEtran}
\usepackage{amsmath,amsfonts,amssymb,amsthm}
\usepackage{bm}
\usepackage{algorithmic}
\usepackage{algorithm}
\usepackage{array}
\usepackage{booktabs}
\usepackage{multirow}
\usepackage[caption=false,font=normalsize,labelfont=sf,textfont=sf]{subfig}
\usepackage{textcomp}
\usepackage{stfloats}
\usepackage{float}
\usepackage{url}
\usepackage{graphicx}
\usepackage{cite}
\usepackage[hidelinks]{hyperref}
\newtheorem{proposition}{Proposition}

\newcommand{\E}{\mathbb{E}}

\newcommand{\xc}{\bar{x}_c}
\newcommand{\xu}{\bar{x}_\emptyset}
\newcommand{\gc}{g_c}
\newcommand{\ours}{\textsc{(ours)}}

\begin{document}

\title{Guidance at Pure Noise Erodes Speaker Similarity in Few-Step Token-Conditioned Flow-Matching TTS}

\author{Jaeho Lee and Han-Lim Choi%
\thanks{This work has been submitted to the IEEE for possible publication. Copyright may be transferred without notice, after which this version may no longer be accessible.}%
\thanks{Jaeho Lee is with the Department of Aerospace Engineering,
        KAIST, Daejeon, Republic of Korea.
        {\tt\small 16jhlee@kaist.ac.kr}}%
\thanks{Han-Lim Choi is with the Department of Aerospace Engineering,
        KAIST, Daejeon, Republic of Korea.
        {\tt\small hanlimc@kaist.ac.kr}}%
}

\markboth{Preprint submitted to IEEE Transactions on Audio, Speech and Language Processing}%
{Lee and Choi: Guidance at Pure Noise Erodes Speaker Similarity}

\maketitle

\begin{abstract}
Text-to-speech systems built from a language model and a flow-matching mel decoder now run the decoder with two to four Euler steps instead of the official ten or more. At two steps the word error rate (WER) is unchanged, but speaker similarity (SIM) drops by one to six points. This paper shows that most of this loss comes from classifier-free guidance on the first step. At pure noise the network can only return the conditional mean, so guidance shifts every trajectory by nearly the same vector. Later steps do not remove this shift, and its size grows with the length of the first step, so it is negligible with ten or more steps and accounts for most of the loss with two. Two training-free samplers are derived from this analysis. First-step unguided sampling (FSU) leaves the first step unguided and recovers $84$--$91\%$ of the two-step SIM loss on CosyVoice~2/3 with no significant WER increase. FSU-Mid replaces the two Euler steps by one midpoint step with an unguided first evaluation, recovers $77$--$85\%$ on IndexTTS-2 and matches the ten-step teacher on CosyVoice~2 and GLM-TTS. Both use three network function evaluations, one fewer than the official two-step sampler, and FSU-Mid gives the highest mean recovery among eight published guidance rules re-run at two steps on four decoders.
\end{abstract}


%=====================================================================
\section{Introduction}
\label{sec:intro}
\begin{figure}[t]
\centering
\includegraphics[width=\columnwidth]{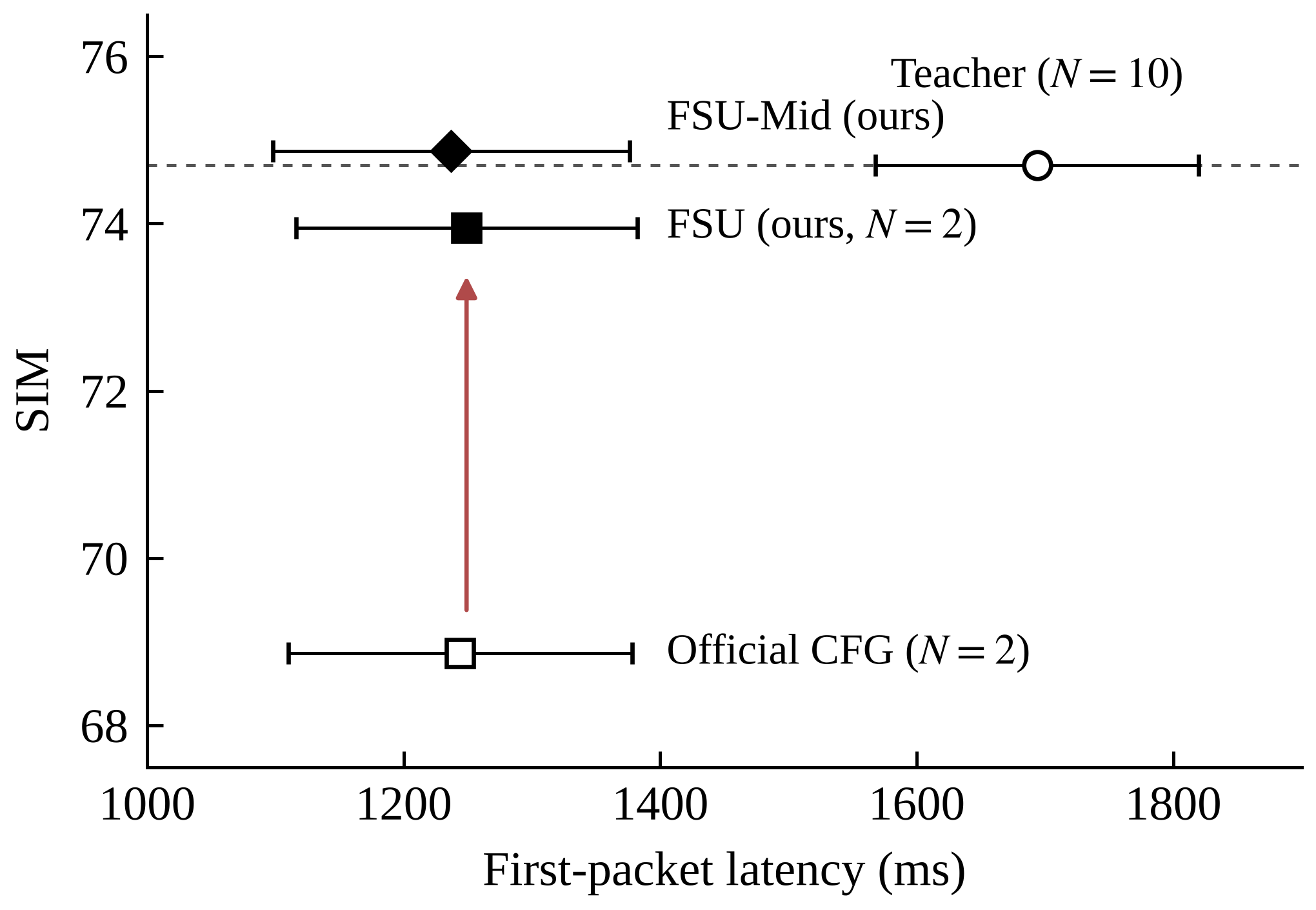}
\caption{First-packet latency and SIM on the official CosyVoice~2 streaming pipeline (test-en, 50 requests). The dashed line marks the ten-step teacher. FSU and FSU-Mid recover most of the two-step loss at the same latency.}
\label{fig:intro_latency}
\end{figure}

\IEEEPARstart{M}{ost} zero-shot text-to-speech (TTS) systems now share a two-stage design~\cite{audiolm,audioldm2,instructtts}. An autoregressive language model (LM) emits discrete speech tokens~\cite{hubert,soundstream,valle,cosyvoice1}, and a flow-matching decoder~\cite{lipman,rectflow} turns the tokens and a speaker prompt into a mel-spectrogram with an Euler solver and classifier-free guidance (CFG)~\cite{hosalimans,cosyvoice2,indextts2,glmtts,maskgct}. The official decoders run ten or more steps with two network evaluations per step, and under concurrent requests the decoder is the component that fills the GPU~\cite{vllmomni}. Production systems therefore reduce the decoder to two to four evaluations, typically through distillation~\cite{progdist,consistency,meanflow,dotstts,flashtts,zipvoice}. Fig.~\ref{fig:intro_latency} shows the trade-off on CosyVoice~2 together with first-step unguided (FSU) sampling, the rule proposed in this paper. Going from ten to two Euler steps shortens the first audio packet by $27\%$ but lowers speaker similarity (SIM) by about six points while word error rate (WER) stays unchanged, and FSU recovers most of this loss at the same latency. Every few-step system surveyed reports the same loss~\cite{dotstts,flashtts,zipvoice,intmeanflow,pci} and accepts it as a cost of acceleration, and no published work explains its cause.

That early guidance can be harmful is known in image synthesis. Restricting guidance to an interval of noise levels~\cite{kynk}, lowering it early in sampling~\cite{stagewise,wangsched}, rescaling, projecting or damping it~\cite{cfgzero,mambog,apg,cfgpp}, replacing the unconditional branch~\cite{autoguidance} and annealing the condition~\cite{cads} all improve sample quality, and analyses of simple models show that guidance at high noise does not sample the intended distribution~\cite{predcorr,chidambaram}. These works operate at five to 250 steps and explain the gain through mode dropping~\cite{kynk} or norm growth~\cite{stagewise}. None of them isolates the guidance of a single step, relates its effect to the step length, or studies speech. In speech, Selective-CFG~\cite{selcfg}, the ZipVoice solver~\cite{zipvoice} and JRR-CFG~\cite{jrr} vary guidance over time at ten or more steps, and all keep it on the first step, where conditioning is commonly thought to be established~\cite{adaptiveguidance,howmuch}.

\begin{figure*}[t]
\centering
\includegraphics[width=0.46\textwidth]{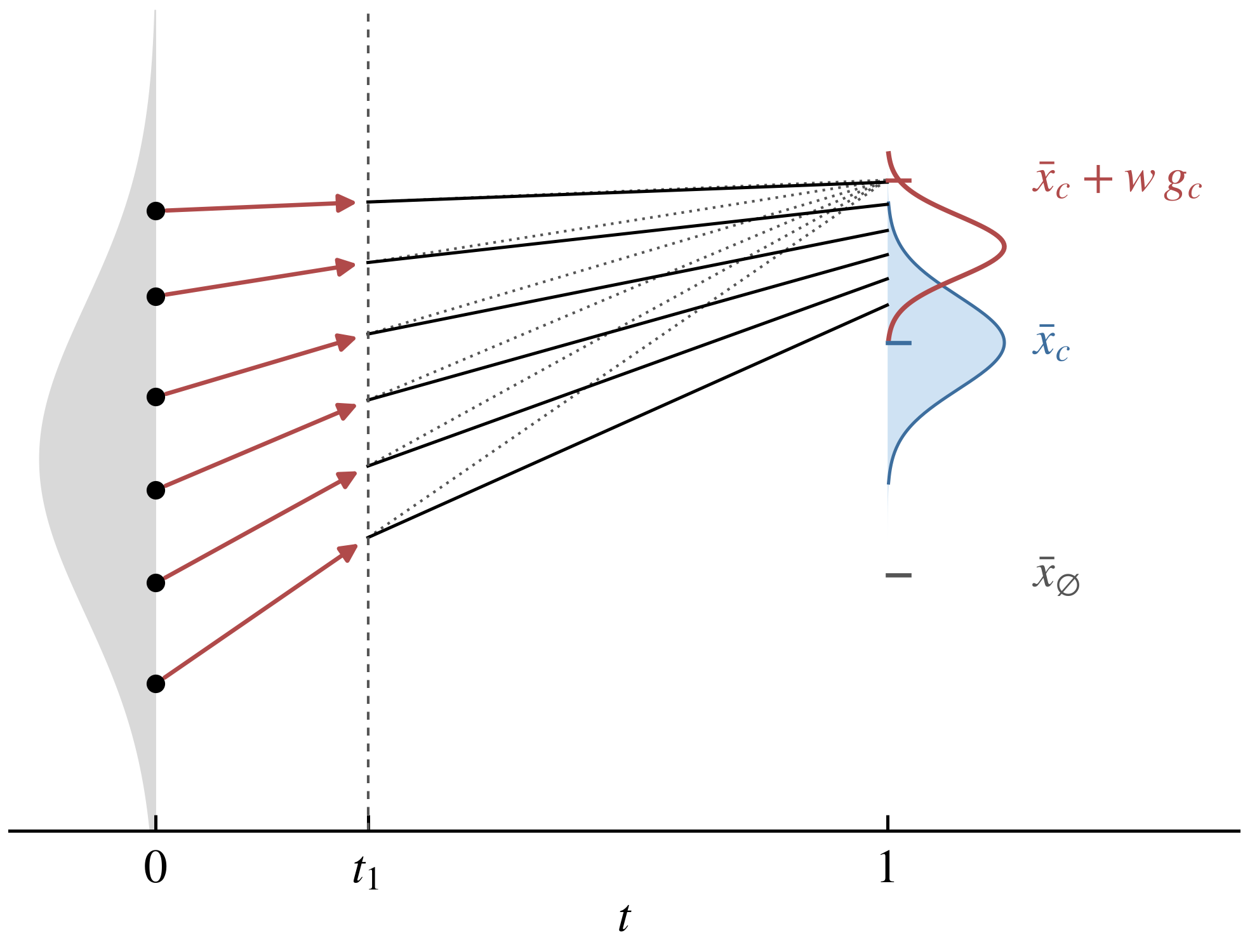}
\hfil
\includegraphics[width=0.46\textwidth]{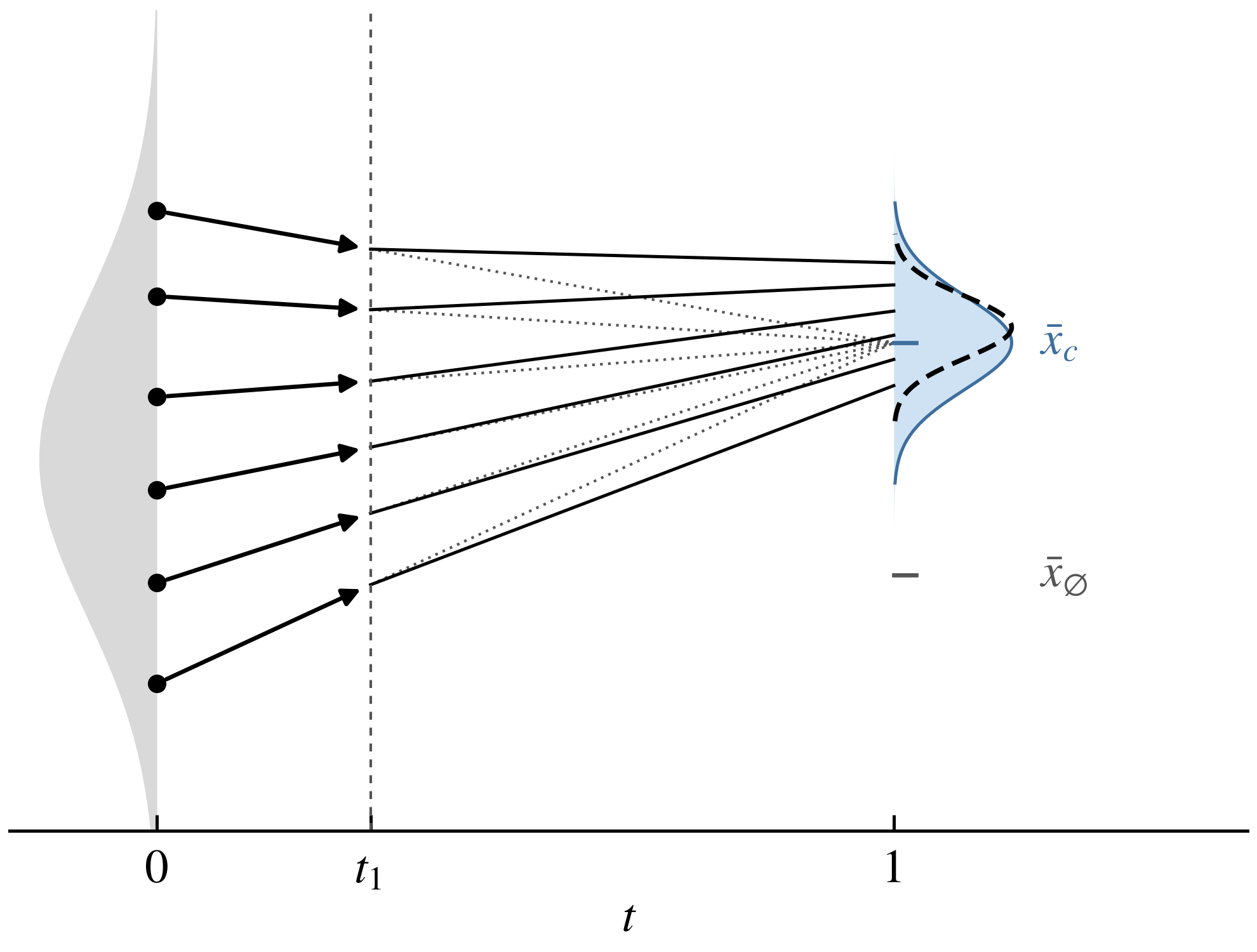}
\caption{Two Euler steps from noise ($t=0$) to data ($t=1$) in one dimension. Left, a guided first step sends every trajectory toward $\bar x_c+w\,g_c$, beyond the conditional mean $\bar x_c$, and shifts the output (red) off the conditional distribution (blue). Right, with FSU the trajectories head for $\bar x_c$ and the output (dashed) stays on it. $g_c=\bar x_c-\bar x_\emptyset$.}
\label{fig:intro_concept}
\end{figure*}

This paper traces the few-step SIM loss of token-conditioned decoders to CFG on the step that starts at pure noise. At that point the network can only predict the average output for the given tokens and prompt, so guidance pushes every trajectory past this average by nearly the same amount (Fig.~\ref{fig:intro_concept}). The distribution of outputs over noise seeds is therefore shifted rather than blurred, and later steps do not remove the shift. Because the shift grows with the length of the first step, it is negligible with tens of steps and accounts for most of the loss with two. FSU leaves that step unguided and removes the shift with no training or tuning and with one network evaluation fewer. The analysis is stated for the state after the first step, and its consequences for the output are measured on four token-conditioned decoders.

The evidence comes from four token-conditioned decoders and two text-conditioned controls. A direct test of the one-step closed form in mel space and controls that switch guidance on and off step by step attribute the loss to the first step, since guiding that step alone is worse than using no guidance at all. FSU recovers $84$--$91\%$ of the two-step SIM loss on CosyVoice~2/3 without a significant WER increase on Seed-TTS-eval. Its midpoint variant FSU-Mid recovers $77$--$85\%$ on IndexTTS-2 and matches or exceeds the ten-step teacher on CosyVoice~2 and GLM-TTS. Both use three network evaluations, against twenty for the teacher. The analysis also delimits where the rule applies. The loss is large when the unconditional branch erases the speech tokens, whereas in text-conditioned decoders first-step guidance establishes the alignment and should be kept. Eight published guidance rules re-evaluated at two steps on the same four decoders recover more SIM the more first-step guidance they remove, and FSU-Mid gives the highest mean recovery among them.

At two steps the first step covers the first $29\%$ of the trajectory, so FSU coincides with the high-noise end of the guidance interval of~\cite{kynk}. The analysis explains why that end carries the whole few-step effect, why the effect appears only in token-conditioned decoders, and why the low-noise end, which costs speech $0.2$--$0.9$ SIM at ten steps, should not be adopted. No distilled token-conditioned decoder has public weights, so the experiments run the public teachers with few Euler steps, and distilled students are left to future work. Official implementations evaluate the conditional and unconditional branches in one batch, so removing one branch evaluation saves computation but leaves the wall-clock time nearly unchanged, and FSU improves quality at the same latency.

%=====================================================================
\section{Background}
\label{sec:background}

\subsection{Flow-Matching Decoders and Classifier-Free Guidance}
A flow-matching decoder~\cite{lipman,rectflow,voicebox,matcha} maps Gaussian noise $\epsilon$ to a mel-spectrogram $x$ that is consistent with a conditioning $c=(\tau,s)$, where $\tau$ denotes the speech tokens and $s$ the speaker part, a speaker embedding together with a prompt mel. Time $t\in[0,1]$ runs from noise ($t=0$) to data ($t=1$), the state on the linear path is $z_t=tx+(1-t)\epsilon$, and the network learns the conditional velocity $u^c_t(z)=\E[x-\epsilon\mid z_t=z,c]$. Sampling integrates $\dot z_t=u^c_t(z_t)$ from $z_0=\epsilon$ to $t=1$. Classifier-free guidance~\cite{hosalimans} also evaluates an unconditional velocity $u^\emptyset_t$, obtained by erasing part of $c$, and follows
\begin{equation}
\tilde u_t=u^c_t+w\,(u^c_t-u^\emptyset_t),
\label{eq:cfg}
\end{equation}
where $w$ is the guidance weight ($w=0.7$ in CosyVoice~2/3 and GLM-TTS). Papers that write $\tilde u=u^\emptyset+g\,(u^c-u^\emptyset)$ use the guidance scale $g=1+w$, and every external rule in this paper is converted to $w$.

Decoders differ in what the unconditional branch erases. The all-null branch erases the tokens, the speaker embedding and the prompt mel, the speaker-null branch erases only the speaker part, and the token-null branch erases only the tokens. Text-conditioned decoders use a text-null branch, which erases the text and keeps the speech prompt.

The sampler takes $N$ Euler steps on a grid $0=t_0<t_1<\dots<t_N=1$,
\begin{equation}
z_{t_k}=z_{t_{k-1}}+h_k\,\tilde u_{t_{k-1}}(z_{t_{k-1}}),\qquad h_k=t_k-t_{k-1},
\end{equation}
with weight $w_k$ in step $k$. The first step has length $\Delta t_1=t_1$ and weight $w_1$. CosyVoice and GLM-TTS use the cosine grid $t_k=1-\cos(\frac{\pi}{2}\frac kN)$ and IndexTTS-2 a uniform grid. Cost is counted in network function evaluations (NFEs), two for a guided step and one for an unguided step.

\subsection{Few-Step Decoding in Deployment}
\label{sec:why}
Streaming TTS is judged by the delay before the first audio packet, and the decoder is on the critical path of every chunk. A guided Euler step costs two network evaluations, so the decoder cost grows linearly with the number of steps. The cost matters most when the LM runs on a server and the decoder and vocoder run on the user's device, since the time of each step then depends on that device rather than on server hardware. Production systems therefore target two to four network evaluations~\cite{dotstts,flashtts,zipvoice,pci,intmeanflow}, and the SIM loss described in Section~\ref{sec:intro} appears in exactly this range. On the official CosyVoice~2 streaming pipeline, the decoder takes $37.7\%$ of GPU time at ten steps and $19.4\%$ at two, which shortens the first packet by $27\%$ and raises throughput by $1.21\times$ at a concurrency of one and by $1.44\times$ at sixteen. In the unaccelerated setup measured here the LM accounts for most of the remaining latency, and the decoder share grows again once the LM is accelerated~\cite{vllmomni}. Table~\ref{tab:latency} in Appendix~\ref{app:A} gives the full measurements, including the cost of FSU. The analysis holds for any step count. One step is the limiting case, in which the first step spans the whole trajectory.

\subsection{Guidance Rules and Few-Step TTS}
\label{sec:related}
Guidance intervals~\cite{kynk}, stage-wise CFG~\cite{stagewise}, CFG-Zero$^*$~\cite{cfgzero}, MAMBO-G~\cite{mambog} and terminal-fitted repair~\cite{tfr} reshape guidance over time in image synthesis and explain their gains by mode dropping or norm growth. Proposition~\ref{thm:translation} of Section~\ref{sec:analysis} is elementary and implicit in Gaussian analyses~\cite{stagewise,predcorr}. The contribution here lies in its use. It identifies a nearly seed-invariant displacement with unchanged spread, is tested in mel space, and is tied to the step length and the conditioning branch. In speech, Selective-CFG~\cite{selcfg}, the ZipVoice solver~\cite{zipvoice} and JRR-CFG~\cite{jrr} operate at ten or more steps and keep first-step guidance. Distilled decoders~\cite{dotstts,flashtts,zipvoice,intmeanflow} report the SIM loss without identifying its cause, and prompt-consistency inference~\cite{pci} addresses prompt drift. DMOSpeech~\cite{dmospeech,dmospeech2} reports F0 mode shrinkage under distribution matching, which is the observation closest to the output geometry studied in Section~\ref{sec:geometry}. Table~\ref{tab:rules} re-runs these rules at two steps.

%=====================================================================
\section{Analysis of First-Step Guidance}
\label{sec:analysis}

\subsection{Localizing the Few-Step Loss}
\label{sec:tax}
\begin{figure*}[t]
\centering
\includegraphics[width=0.92\textwidth]{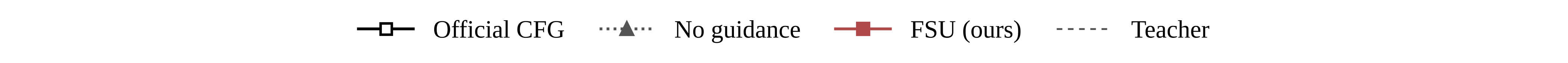}\\[-2pt]
\includegraphics[width=0.32\textwidth]{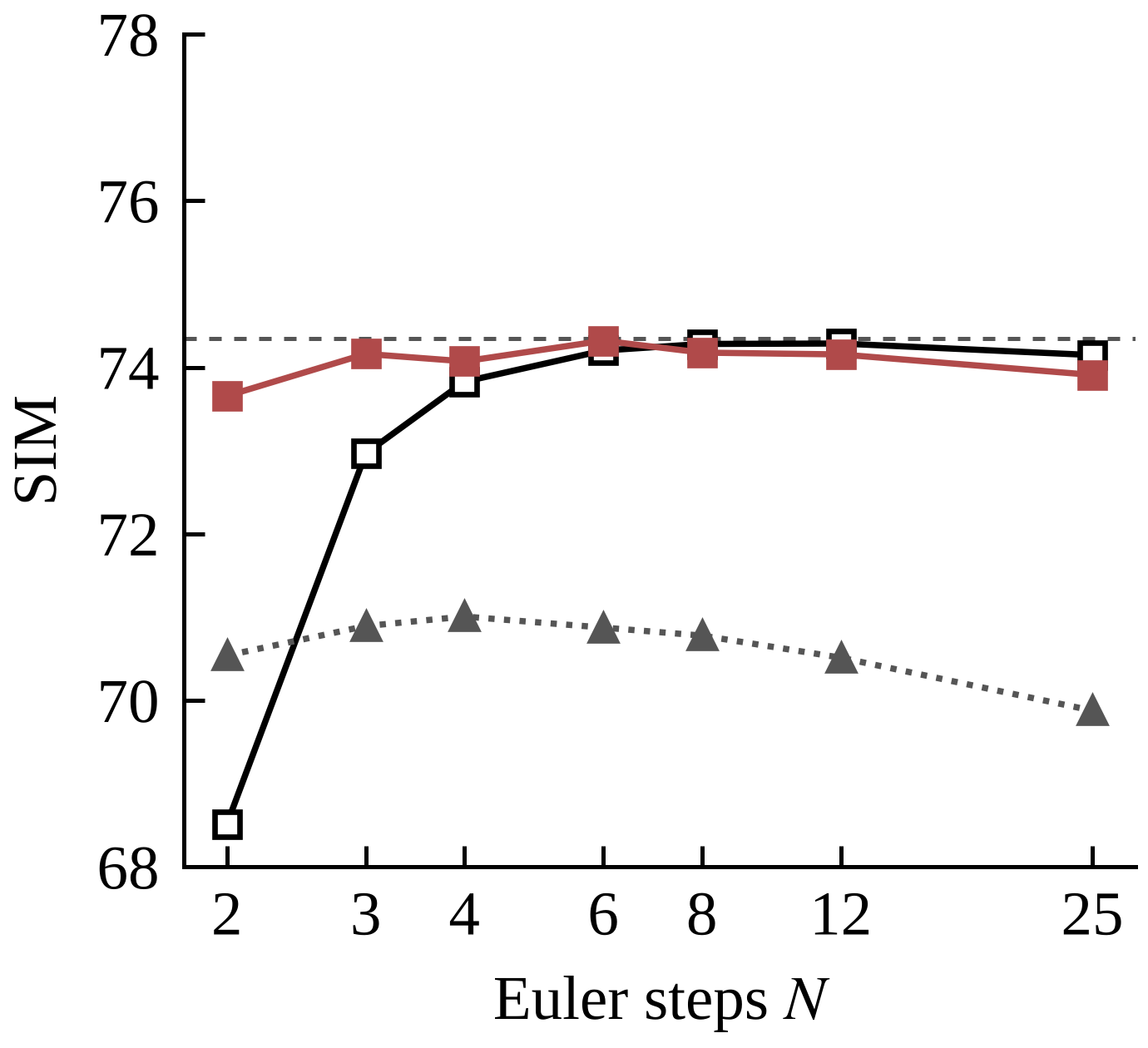}
\hfil
\includegraphics[width=0.32\textwidth]{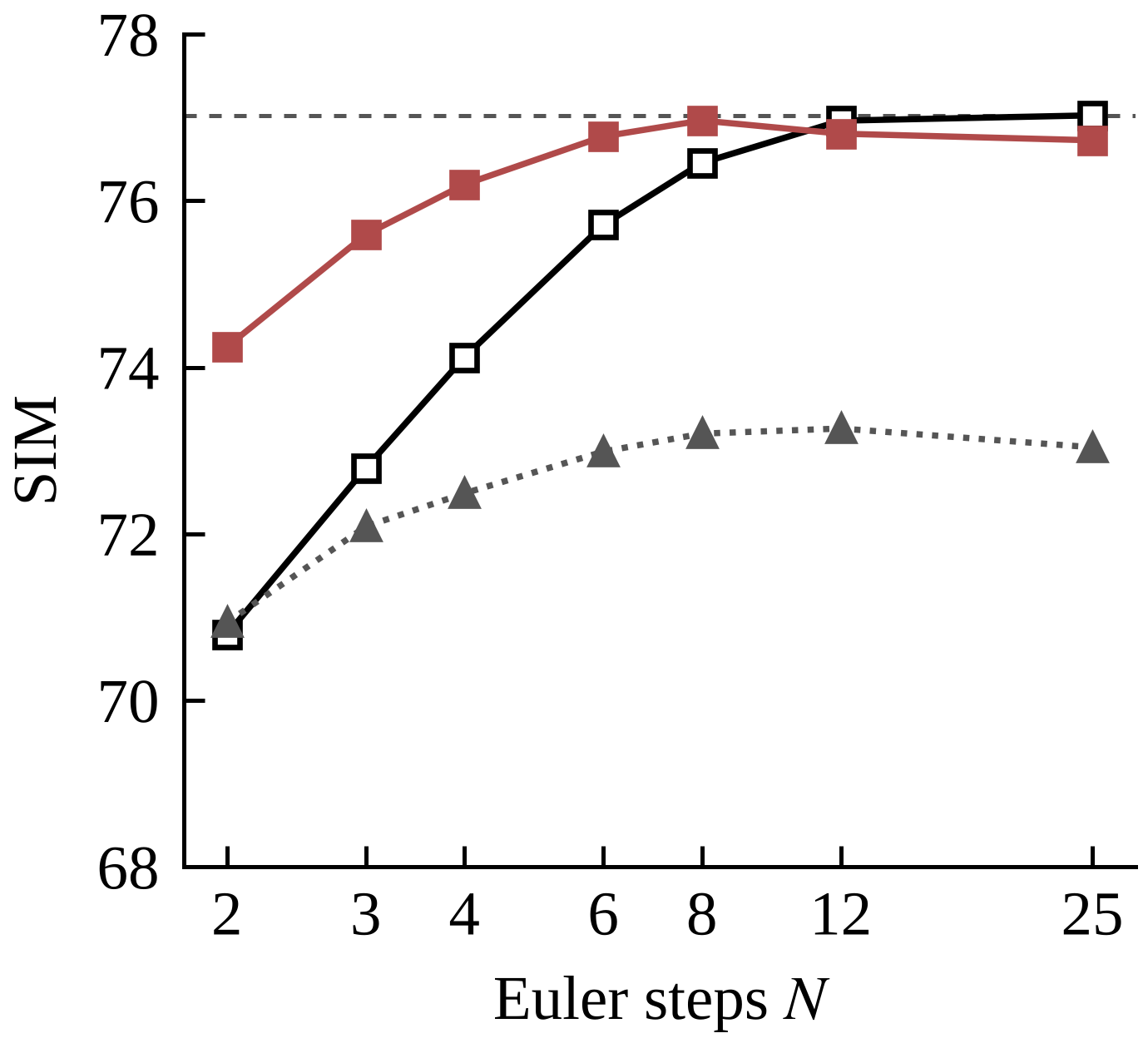}
\hfil
\includegraphics[width=0.32\textwidth]{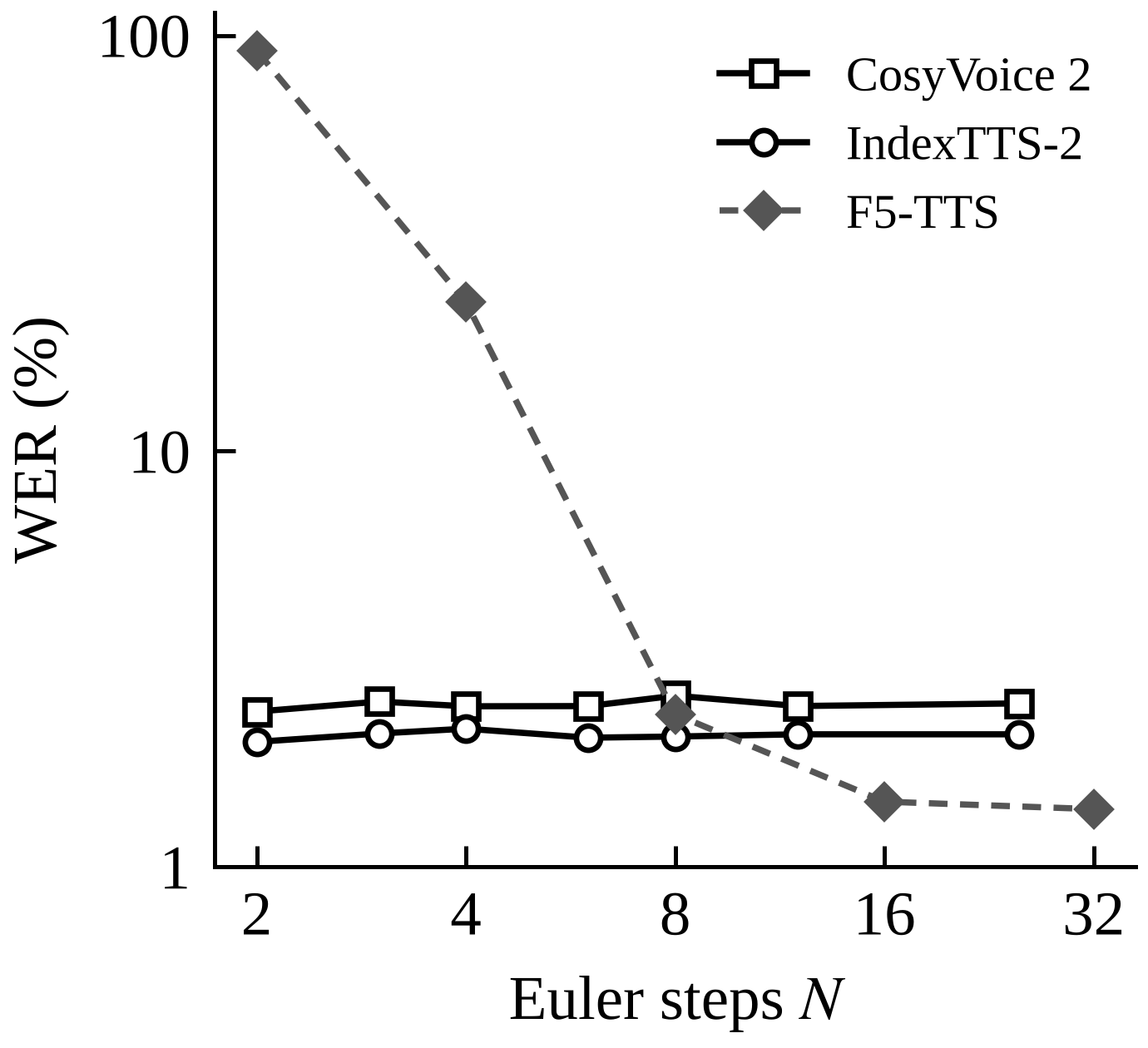}
\caption{SIM and WER against the number of Euler steps (test-en, analysis subset). Left and middle, SIM of CosyVoice~2 and IndexTTS-2 with the official CFG, without guidance and with FSU, and the teacher (dashed). Right, WER with the official CFG, including the text-conditioned F5-TTS.}
\label{fig:obs}
\end{figure*}

Fig.~\ref{fig:obs} reduces the number of Euler steps of two token-conditioned decoders from different groups, CosyVoice~2 and IndexTTS-2, and compares three samplers on each. With the official guidance, CosyVoice~2 loses only $0.5$ SIM points at four steps, then $1.4$ at three and $5.8$ at two, and IndexTTS-2 loses $2.9$ points at four and $6.2$ at two. WER stays between $1.9$ and $2.7\%$ at every step count (Fig.~\ref{fig:obs}, right). The tokens fix the content and the rhythm, so the speaker is the only unconstrained attribute. In the text-conditioned F5-TTS, whose content is not fixed in advance, WER rises instead.

The two other samplers show where the loss comes from. Turning guidance off on every step does not help, and its SIM stays three to six points below the teacher at every step count, so guidance itself is needed. Turning guidance off on the first step only, which is FSU, recovers most of the loss. At two steps it raises SIM from $68.5$ to $73.7$ on CosyVoice~2 and from $70.8$ to $74.2$ on IndexTTS-2, and it reaches the teacher with about half as many steps as the official sampler. The loss is therefore caused neither by guidance as such nor by the small number of steps alone, but by guidance on the first step, the step that starts at pure noise.

\subsection{First-Step Translation}
\label{sec:translation}
Write $\xc=\E[x\mid c]$ and $\xu=\E[x\mid\emptyset]$ for the conditional and unconditional means, $\gc=\xc-\xu$ for the guidance direction, and $r_c^2=\E[\|x-\xc\|^2\mid c]$ for the spread of the conditional distribution. The first Euler step starts at $z_0=\epsilon$, which carries no information about the output. The network can then return only the average output for the given tokens and prompt, and guidance pushes every trajectory away from the unconditional average by the same vector (Fig.~\ref{fig:intro_concept}). The following proposition formalizes this for a network that returns the exact velocities.

\begin{proposition}[First-step translation]\label{thm:translation}
If the network returns the exact velocities $u^c_t$ and $u^\emptyset_t$, the state after a first Euler step of length $t_1$ with weight $w_1$ is
\begin{equation}
z^{w_1}_{t_1}(\epsilon)=z^{0}_{t_1}(\epsilon)+t_1w_1\gc
\label{eq:translation}
\end{equation}
for every noise sample $\epsilon$. Its root-mean-square distance from the conditional paths $(1-t)\epsilon+tx$, $x\sim p(x\mid c)$, through the same $\epsilon$ is
\begin{equation}
D(w_1)=t_1\sqrt{r_c^2+w_1^2\|\gc\|^2}.
\label{eq:dist}
\end{equation}
\end{proposition}
\begin{proof}
Since $z_0=\epsilon$ is independent of $(x,c)$,
\begin{equation}
u^c_0(\epsilon)=\E[x-\epsilon\mid z_0=\epsilon,c]=\xc-\epsilon,\qquad u^\emptyset_0(\epsilon)=\xu-\epsilon,
\label{eq:noisevel}
\end{equation}
so $u^c_0-u^\emptyset_0=\gc$ for every $\epsilon$. The Euler update gives $z^{w_1}_{t_1}=\epsilon+t_1(\xc-\epsilon+w_1\gc)$, which is \eqref{eq:translation}. The difference from the conditional path is $t_1(\xc+w_1\gc-x)$, and its mean square is $t_1^2(r_c^2+w_1^2\|\gc\|^2)$ because $\E[x-\xc\mid c]=0$ removes the cross term.
\end{proof}

The proposition establishes four facts about the first-step state. First, guidance adds the same vector to every trajectory, so it moves the whole distribution of $z_{t_1}$ without changing its spread, a displacement rather than a blur. Second, the size of this displacement,
\begin{equation}
E_1=|w_1|\,\|\gc\|\,\Delta t_1,
\label{eq:e1}
\end{equation}
is proportional to the length of the first step. On the cosine grid $\Delta t_1$ is $0.29$ at two steps and $0.01$ at ten, so the same weight moves the first state about twenty-five times farther at two steps, which explains why the loss appears only at low step counts. Third, \eqref{eq:dist} splits the deviation into an averaging term $r_c$, present in any one-step prediction, and a guidance term $|w_1|\,\|\gc\|$, which vanishes exactly at $w_1=0$ and grows for any other weight, positive or negative. Their ratio
\begin{equation}
\kappa_c=r_c/\|\gc\|
\label{eq:kappa}
\end{equation}
gives the share of the deviation that guidance is responsible for. Fourth, the size of $\gc$ depends on what the unconditional branch erases. For nested conditioning $c=(\tau,s)$,
\begin{equation}
\E\|\bar x_{\tau,s}-\xu\|^2=\E\|\bar x_{\tau,s}-\bar x_\tau\|^2+\E\|\bar x_\tau-\xu\|^2,
\label{eq:tower}
\end{equation}
since $\E[\bar x_{\tau,s}-\bar x_\tau\mid\tau]=0$, so an all-null branch adds a token component to $\gc$ that a speaker-null branch does not have.

Three further claims lie outside the proposition and are supported empirically rather than proven. The proposition concerns the state after the first step, and whether the displacement survives to the output depends on the later steps of the trained network. It is expected to survive, as the recovery in Fig.~\ref{fig:obs} suggests, and Section~\ref{sec:persist} measures it directly. The proposition also gives a distance in mel space rather than an effect on the speaker, and the displacement is expected to move the output toward another voice because the tokens leave the speaker as the only free attribute. Finally, removing first-step guidance is expected to help only when the guidance term dominates the averaging term ($\kappa_c<1$). When the averaging term dominates, as in text-conditioned decoders whose first step must still decide the alignment, first-step guidance contributes to the alignment, and removing it should raise WER. The proposition motivates this boundary, and Section~\ref{sec:branch} locates it empirically.

%=====================================================================
\section{First-Step Unguided Sampling}
\label{sec:remedy}
\textbf{FSU.} Section~\ref{sec:translation} suggests the simplest remedy, since the guidance term vanishes exactly when the first step is unguided. FSU evaluates only the conditional network on every step whose start time $t_{k-1}$ lies below a threshold $t^*$ and applies \eqref{eq:cfg} on the remaining steps. FSU needs no training and saves one network evaluation per unguided step.

\textbf{Threshold.} For exact velocities, Proposition~\ref{thm:translation} holds exactly at $t=0$, where the state carries no information about the output. As $t$ grows the state starts to determine the output, and the guidance difference shrinks and begins to depend on the seed. The threshold $t^*$ marks the end of this noise-dominated regime. It is set to $t^*=0.1$, where the guidance magnitude along teacher paths has fallen to about a third of its value at noise in all four decoders (Section~\ref{sec:gamma}). The rule is applied to start times, so the number of unguided steps follows from the grid. On the cosine grid the start times below $0.1$ are $t_0$ at two and three steps ($t_1=0.29$ and $0.13$), $t_0$ and $t_1=0.08$ at four steps and $t_0$, $t_1=0.01$ and $t_2=0.05$ at ten steps, so FSU leaves step~1, steps~1--2 and steps~1--3 unguided. On the uniform grid of IndexTTS-2 it leaves step~1 unguided at two steps and steps~1--3 ($t_2=0.08$) at 25. At two steps every grid gives the same rule, an unguided first step. A ramp $w_k=w\min(1,t_{k-1}/0.2)$ gives the same results within seed noise.

\textbf{FSU-Mid.} FSU removes the guidance term, but the ordinary two-step Euler error remains and can be large, especially on a uniform grid whose first step spans half of the trajectory. FSU-Mid replaces the two Euler steps by one midpoint step. It evaluates $u^c$ at $t=0$ without guidance, takes a half step to $t=0.5$, evaluates \eqref{eq:cfg} there and takes the full step with that velocity. This costs three evaluations, the same as FSU and one fewer than the guided two-step baseline.

\textbf{Applicability.} FSU suits decoders whose unguided SIM barely changes with the number of steps (CosyVoice, GLM-TTS), and FSU-Mid suits decoders with a large truncation error (IndexTTS-2, Appendix~\ref{app:B}). Both target the two- to four-step regime of token-conditioned decoders ($\kappa_c<1$). At one step the first step is the whole trajectory, and the official guidance is kept there (Appendix~\ref{app:C}). A negative first-step weight is not recommended (Section~\ref{sec:controls}). If first-step guidance must be kept, an unconditional branch that keeps the tokens reduces the loss by itself, following \eqref{eq:tower}. This holds for token-conditioned decoders only. In F5-TTS and ZipVoice the branch that erases only the speech prompt retains $0.8$ of the all-null magnitude (Table~\ref{tab:branch}), because without tokens the prompt is what fixes the content.

%=====================================================================
\begin{figure*}[t]
\centering
\includegraphics[width=0.45\textwidth]{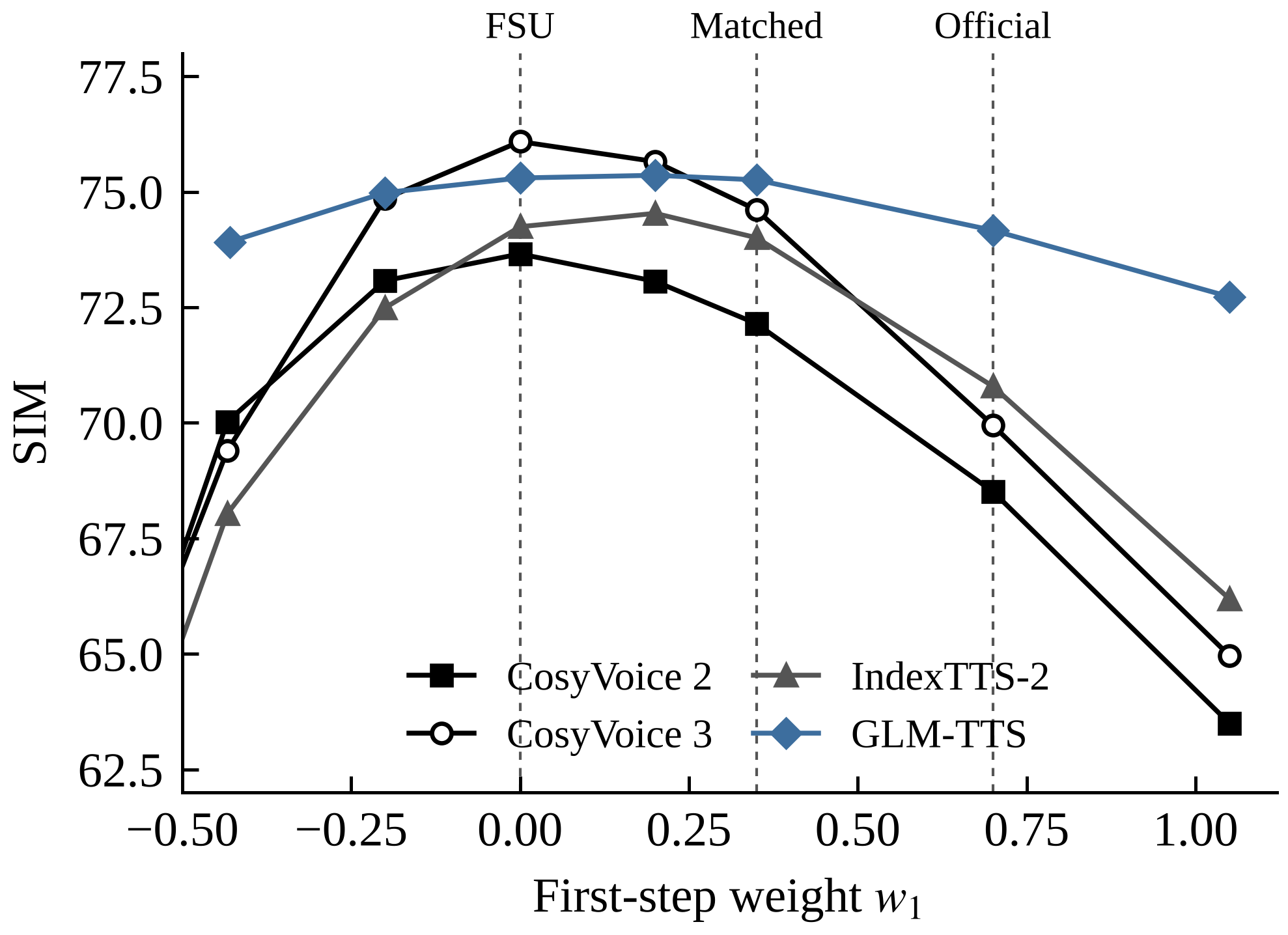}
\hfil
\includegraphics[width=0.45\textwidth]{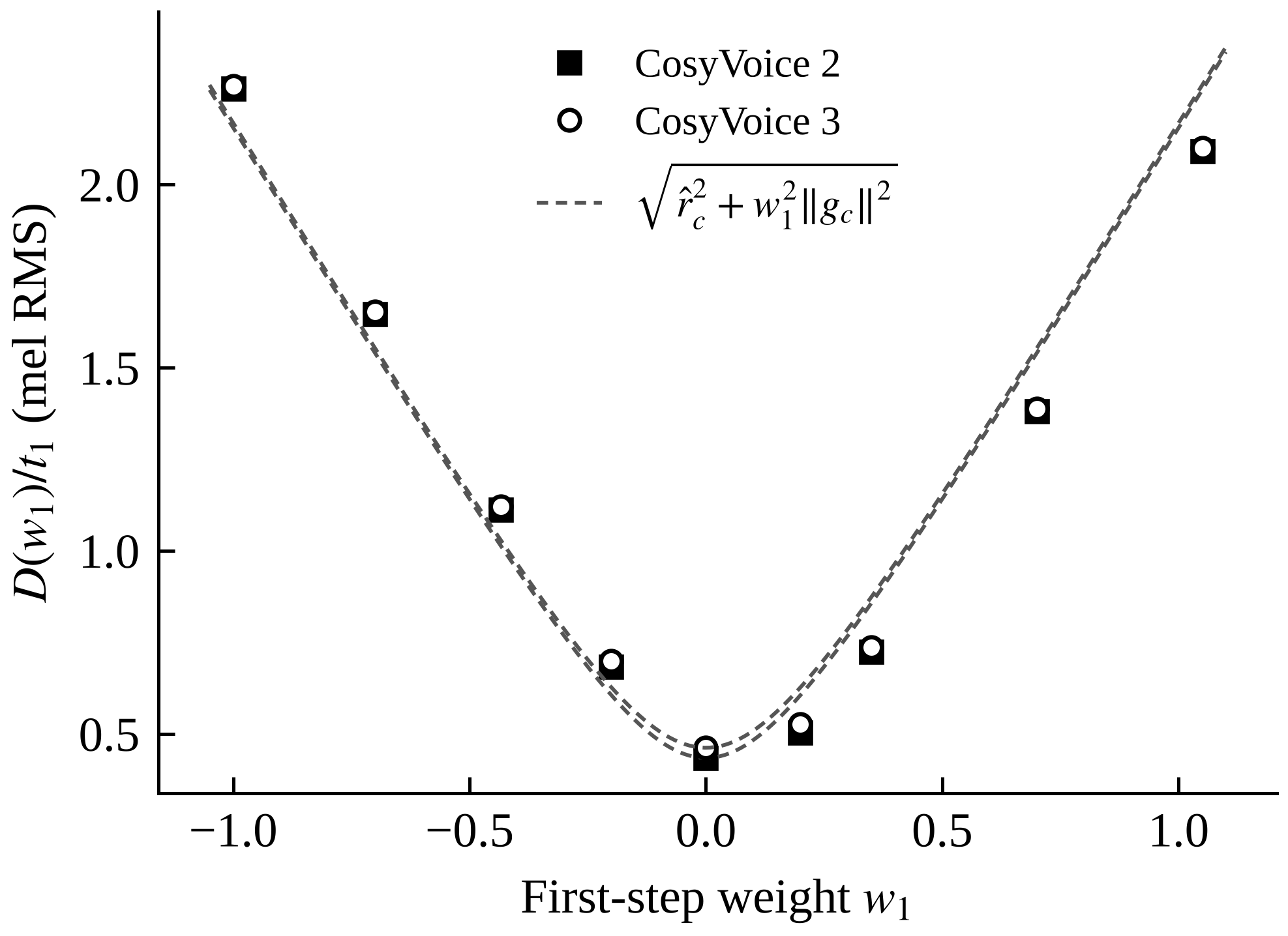}
\caption{First-step weight $w_1$ swept with $w_2=0.7$ (test-en, two steps). Left, SIM of four decoders, with vertical lines at FSU ($0$), the weight matched to the teacher's early guidance ($0.35$) and the official CFG ($0.7$). Right, measured first-step deviation $D(w_1)/t_1$ (markers) and the closed form \eqref{eq:dist} (dashed).}
\label{fig:wsign}
\end{figure*}

\section{Experimental Setup}
\label{sec:protocol}
All experiments use public checkpoints and change only the sampler.\footnote{Code and audio samples are available at \url{https://github.com/16jhlee/FSU}.} Table~\ref{tab:decoders} lists the decoders. The analysis covers four token-conditioned decoders, and the text-conditioned F5-TTS and ZipVoice serve as controls for the boundary of Section~\ref{sec:boundaries}. Every few-step configuration is compared with the official decoder at its official step count, called the teacher.

\begin{table}[t]
\caption{Decoders. Steps is the official step count, used by the teacher. Sway and t-shift are the nonuniform grids of F5-TTS and ZipVoice.}
\label{tab:decoders}
\centering\footnotesize
\setlength{\tabcolsep}{2.5pt}
\begin{tabular}{llccl}
\toprule
Decoder & Condition & Grid & Steps & Uncond.\ branch \\
\midrule
CosyVoice~2~\cite{cosyvoice2} & tokens & cosine & 10 & all-null \\
CosyVoice~3~\cite{cosyvoice3} & tokens & cosine & 10 & all-null \\
IndexTTS-2~\cite{indextts2} & tokens & uniform & 25 & all-null \\
GLM-TTS~\cite{glmtts} & tokens & cosine & 10 & speaker-null \\
F5-TTS~\cite{f5tts} & text & sway & 32 & all-null \\
ZipVoice~\cite{zipvoice} & text & t-shift & 16 & text-null \\
\bottomrule
\end{tabular}
\end{table}

The evaluation sets are Seed-TTS-eval~\cite{seedtts} (test-en 1,088, test-zh 2,020 and hard 400 utterances) and LibriSpeech-PC test-clean~\cite{librispeech,lspc} (1,127 prompt--target pairs), complemented by multilingual, noisy-prompt and streaming sets, and the analysis experiments use a fixed subset of 300/300/100 Seed-TTS-eval utterances. SIM is the cosine similarity ($\times100$) between ERes2Net~\cite{eres2net} speaker embeddings of the output and the prompt, WER (CER for Chinese) is measured with the Seed-TTS-eval ASR~\cite{whisper,paraformer}, and naturalness with five objective predictors~\cite{utmos,utmosv2,dnsmos,nisqa}. Every comparison is paired at the utterance level, with identical LM tokens, length and noise seed. The seed noise $\varepsilon$ of a metric is its utterance-level standard deviation across eight noise seeds of the teacher ($0.42$ for SIM), that is, the variation caused by the noise seed alone. Differences smaller than $\varepsilon$ are referred to as \emph{within seed noise}. Appendix~\ref{app:A} describes the data, metrics, statistical tests and seed noise in full.

\begin{figure*}[t]
\centering
\includegraphics[width=0.45\textwidth]{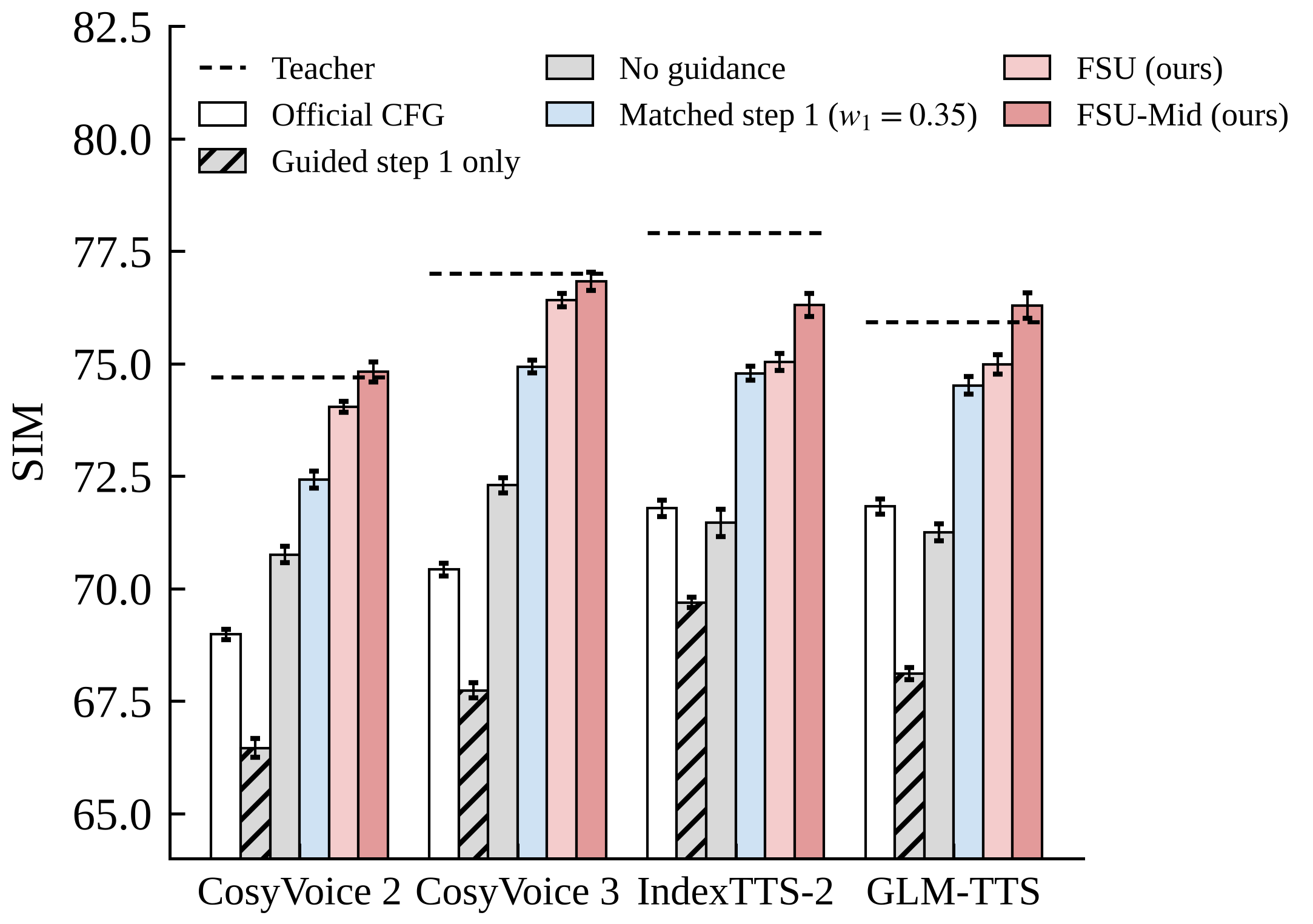}
\hfil
\includegraphics[width=0.45\textwidth]{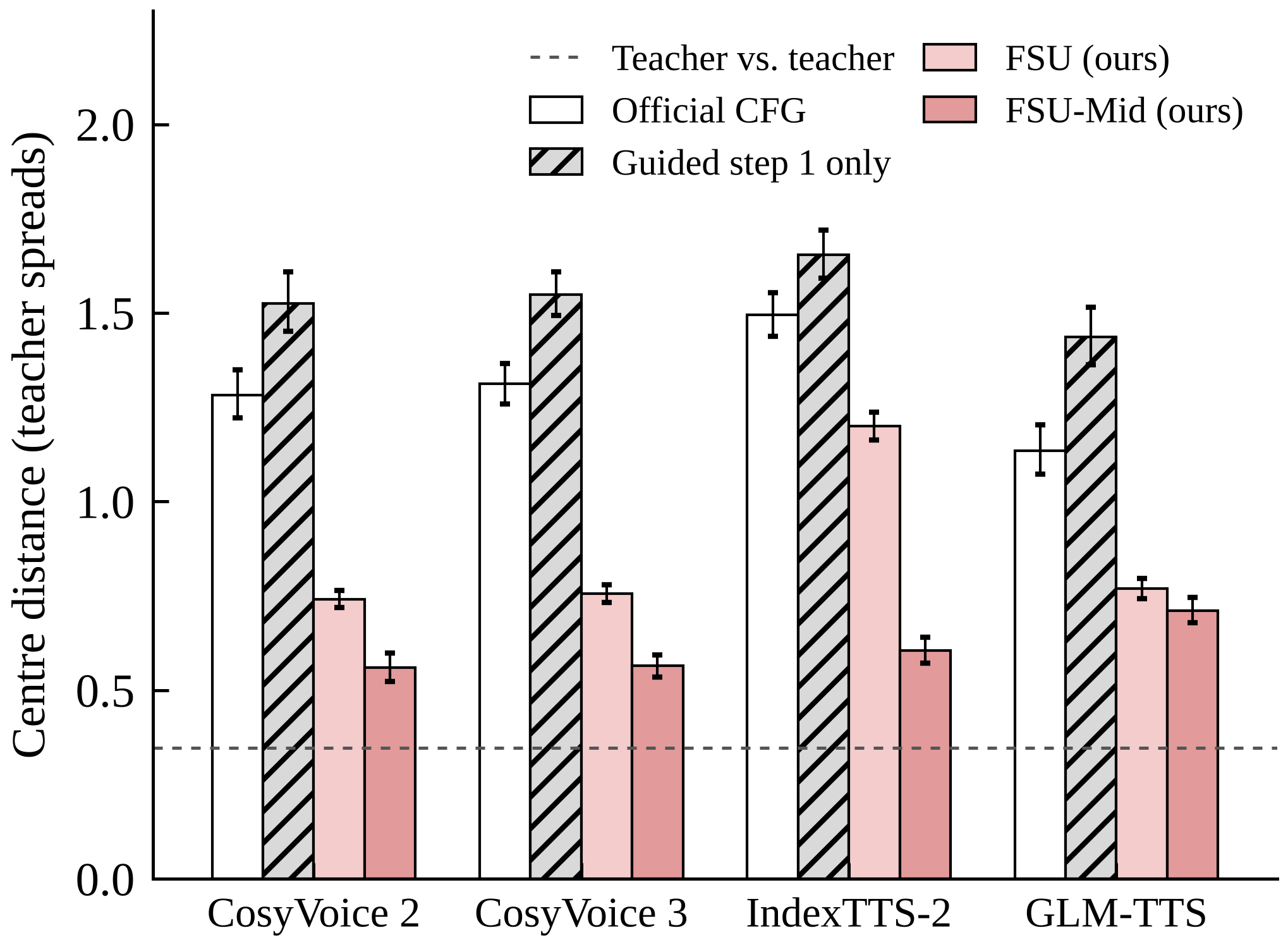}
\caption{Two-step samplers that differ only in where guidance is applied (test-en, 100 utterances, eight seeds, mean and s.d.). Left, SIM, with the teacher as dashed segments. Right, distance between the output and teacher centers, with the teacher-versus-teacher floor (dashed). GLM-TTS uses the all-null branch.}
\label{fig:controls}
\end{figure*}

\begin{figure*}[t]
\centering
\includegraphics[width=0.45\textwidth]{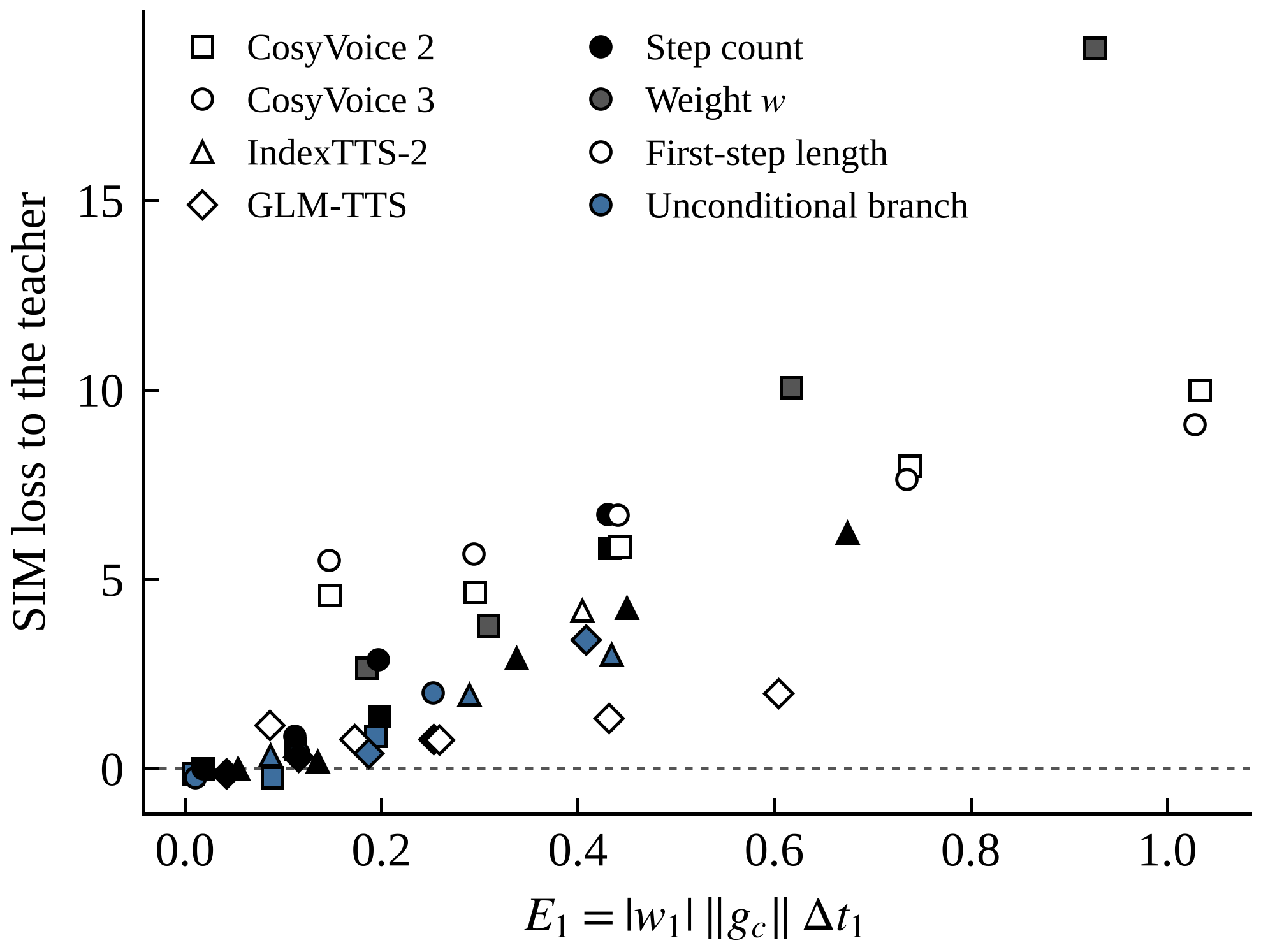}
\hfil
\includegraphics[width=0.45\textwidth]{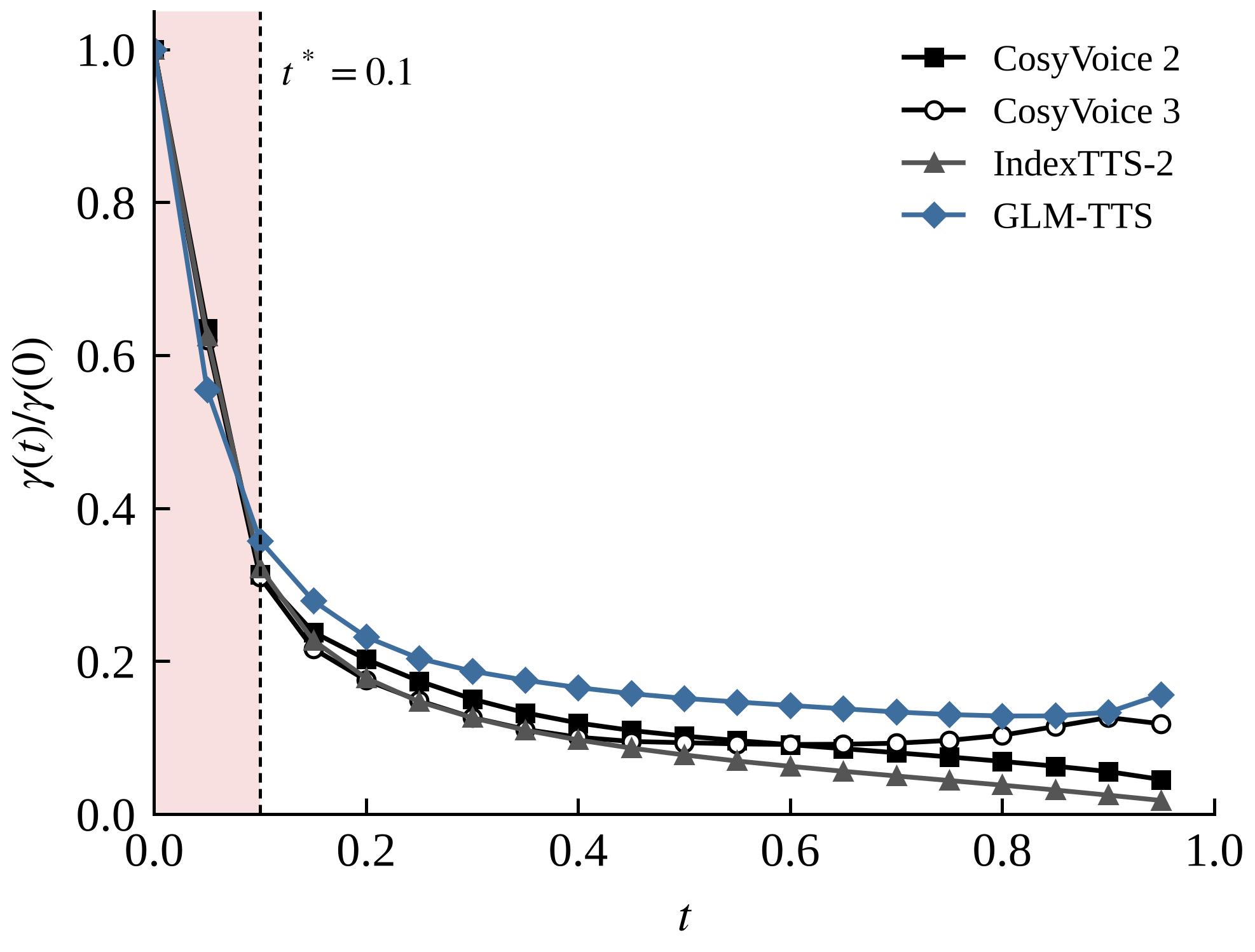}
\caption{Left, SIM loss to the teacher against $E_1$ for every manipulation of step count, weight, first-step length and branch (marker shape gives the decoder, fill the manipulation). Right, guidance magnitude $\gamma(t)$ along teacher trajectories, relative to $t=0$, with the region below $t^*=0.1$ shaded.}
\label{fig:verify}
\end{figure*}

%=====================================================================
\section{Verification of the Analysis}
\label{sec:verify}
This section tests the claims of Section~\ref{sec:translation} in the order they were made. Section~\ref{sec:closedform} tests the assumption of Proposition~\ref{thm:translation}. Section~\ref{sec:geometry} tests the first claim, a translation without change of spread, together with the expected move toward another voice. Sections~\ref{sec:law}, \ref{sec:controls} and~\ref{sec:branch} test the second, third and fourth claims, namely the step-length dependence, the optimum at $w_1=0$ and the effect of the branch, and Section~\ref{sec:branch} also tests the boundary $\kappa_c<1$. Section~\ref{sec:persist} tests the expected persistence of the shift. Appendix~\ref{app:B} gives the measurements behind each subsection.

\subsection{Closed Form in Mel Space}
\label{sec:closedform}
The trained networks satisfy the assumption of Proposition~\ref{thm:translation} closely. At $t=0$ their guidance difference has cosine $0.96$--$0.98$ with an independent estimate of $\gc$ and nearly the same direction for every noise seed, and the measured first-step deviation follows \eqref{eq:dist} with a slope of $0.99\|\gc\|$ in all four decoders (Fig.~\ref{fig:wsign}, right).

\subsection{Output Distribution Geometry}
\label{sec:geometry}
\label{sec:geomafter}
For each utterance, eight noise seeds give a set of speaker embeddings, hereafter a cloud, for the teacher and one for the few-step sampler. The variance ratio compares their spreads, and $\Delta d_{\rm cond}$ is the mean distance of the few-step samples to the teacher center minus the teacher's own leave-one-out distance, which is negative when the cloud is averaged and positive when it is translated. At two steps the official CFG cloud on CosyVoice~2 contracts (variance ratio $0.60$) and moves away from the teacher ($\Delta d_{\rm cond}=+0.043$) while the embedding norm is unchanged. Its center lies $1.3$--$1.8$ teacher spreads from the teacher center on CosyVoice and IndexTTS-2. The shift points toward another voice. Speaker directions are spanned by the outputs for the same tokens under eight different prompts, and on CosyVoice~2 they carry $29\%$ of the energy of the first-step displacement, 24 times the share of the teacher's seed-to-seed difference. FSU returns $\Delta d_{\rm cond}$ to the teacher's level without changing the variance ratio, which is the signature of a translation, and guiding only the first step moves the center farthest (Fig.~\ref{fig:controls}, right). The cloud also remains somewhat narrower than the teacher's, an effect outside the first-step analysis that FSU-Mid reduces (Appendix~\ref{app:B}).

\subsection{Step-Length Dependence}
\label{sec:law}
The left panel of Fig.~\ref{fig:verify} places every manipulation that changes $E_1$ on one axis, namely the step count, the weight, the first-step length and the unconditional branch. Within each decoder the loss is monotone in $E_1$ (Spearman $0.75$--$0.98$), and it vanishes as $E_1\to0$. In these manipulations the total guidance over all steps co-varies with $E_1$, and Section~\ref{sec:controls} separates the two.

\subsection{First-Step Attribution}
\label{sec:controls}
\label{sec:gamma}
\label{sec:twocomp}
On all four decoders the SIM at $w_1=0$ is the maximum or within seed noise of it (Fig.~\ref{fig:wsign}, left). Negative weights, which extrapolate toward the unconditional mean, damage SIM more than positive ones although the mel deviation is symmetric. Guiding the first step alone is worse than using no guidance at all on all four decoders and in each of the eight seeds (on GLM-TTS by $3.1/1.9$ SIM, en/zh). A first step matched to the teacher's early guidance also stays below FSU in all eight seeds (Fig.~\ref{fig:controls}, left). The loss therefore follows guidance applied from pure noise, not the total amount. FSU-Mid lies above FSU on every decoder and in every seed and reaches the teacher's level on CosyVoice and GLM-TTS. The guidance magnitude falls to a third of its value at noise by $t=0.1$ in all four decoders (Fig.~\ref{fig:verify}, right), which sets the threshold $t^*$, and the residual left by FSU is the ordinary truncation error that FSU-Mid targets.

\subsection{Unconditional Branch}
\label{sec:branch}
Erasing the tokens enlarges $\|\gc\|$ by $1.6$--$2.2$ times and moves the branch points in the left panel of Fig.~\ref{fig:verify} to larger $E_1$, and on CosyVoice~2 the all-null branch loses $5.8$ SIM at two steps against $0.9$ for the speaker-null branch. On GLM-TTS, whose released branch is speaker-null, switching the same network to all-null raises the loss from $0.4/0.9$ to $2.8/2.7$ points ($3.4$ on the hard set), FSU recovers $95/79\%$ of it ($70\%$ on hard), and FSU-Mid exceeds the teacher on all three sets. The ratio $\kappa_c$ is $0.21$--$0.49$ for the token-conditioned decoders and $1.09$--$1.97$ for the text-conditioned ones, where removing first-step guidance raises WER (Section~\ref{sec:boundaries}).

\subsection{Persistence Across Steps}
\label{sec:persist}
With the noise fixed, following a guided and an unguided first step with $1$ to $15$ identical later steps leaves a final mel distance of $1.17$--$1.24$ on CosyVoice~2, so the later steps do not remove the shift. Persistence is thus an empirical property of the trained networks, and the loss vanishes with many steps only because $\Delta t_1\to0$.

\begin{table*}[t]
\caption{Published guidance rules at two steps on four decoders (analysis subset, seed 0). Cells give the recovery in percent, en / zh, and Mean averages the eight cells. GLM-TTS uses the all-null branch and its own all-null teacher. Bold marks the best few-step value, and coll.\ marks SIM below 60 or WER above $30\%$. $^\star$Identical to FSU at two steps. $^\dagger$Normalization of the released code corrected. n/a, undefined for IndexTTS-2.}
\label{tab:rules}
\centering\small
\setlength{\tabcolsep}{5pt}\begin{tabular}{lcccccrr}
\toprule
Method & NFE & CosyVoice~2 & CosyVoice~3 & IndexTTS-2 & GLM-TTS & Mean & $\Delta$UTMOS \\
\midrule
Official CFG & 4 & 0 / 0 & 0 / 0 & 0 / 0 & 0 / 0 & 0 & -- \\
\midrule
Guidance interval~\cite{kynk}$^\star$ & 3 & 88 / 85 & 91 / 82 & 55 / 57 & 78 / 69 & 76 & +0.13 \\
Selective-CFG (default)~\cite{selcfg} & 4 & 4 / 16 & 2 / 6 & $-$9 / $-$9 & 2 / $-$6 & 1 & $-$0.05 \\
Selective-CFG (input-text)~\cite{selcfg} & 4 & 85 / 86 & 70 / 56 & 52 / 51 & 63 / 52 & 64 & $-$0.06 \\
CFG-Zero$^*$ scale~\cite{cfgzero} & 4 & 9 / 5 & 3 / $-$2 & 8 / 2 & $-$5 / $-$3 & 2 & +0.04 \\
CFG-Zero$^*$ zero-init~\cite{cfgzero} & 2 & coll. / coll. & coll. / coll. & coll. / coll. & coll. / coll. & coll. & $-$2.05 \\
TV-CFG (released)~\cite{stagewise} & 4 & coll. / coll. & coll. / coll. & coll. / coll. & coll. / coll. & coll. & $-$2.12 \\
TV-CFG (corrected)~\cite{stagewise}$^\dagger$ & 4 & 74 / 65 & 73 / 71 & \textbf{82} / 72 & 61 / 56 & 69 & +0.01 \\
TV-CFG step-up~\cite{stagewise} & 4 & 26 / 18 & $-$8 / 16 & $-$44 / 3 & $-$15 / $-$8 & $-$1 & $-$0.28 \\
MAMBO-G~\cite{mambog} & 4 & 70 / 65 & 73 / 66 & 36 / 39 & 58 / 46 & 57 & +0.25 \\
Terminal-fitted repair~\cite{tfr} & 4 & 35 / 32 & 30 / 32 & 2 / 22 & $-$11 / $-$28 & 14 & +0.15 \\
JRR-CFG~\cite{jrr} & 8 & $-$18 / $-$8 & $-$17 / $-$17 & $-$7 / 0 & $-$17 / $-$2 & $-$11 & $-$0.34 \\
Prompt-consistency~\cite{pci} & 4 & 14 / 13 & 0 / 2 & n/a / n/a & $-$9 / $-$6 & 2 & +0.01 \\
\midrule
\textbf{FSU} \ours & 3 & 88 / 85 & 91 / 82 & 55 / 57 & 78 / 69 & 76 & +0.13 \\
\textbf{FSU-Mid} \ours & 3 & \textbf{104} / \textbf{109} & \textbf{99} / \textbf{96} & 77 / \textbf{79} & \textbf{112} / \textbf{123} & \textbf{100} & $-$0.10 \\
\bottomrule
\end{tabular}
\end{table*}

%=====================================================================
\section{Results}
\label{sec:experiments}

\subsection{Comparison with Guidance Rules}
\label{sec:rules}
Table~\ref{tab:rules} compares FSU and FSU-Mid with the published inference-time guidance rules discussed in Section~\ref{sec:related}, all run at two Euler steps on the four decoders under the same protocol and with the hyperparameters used on CosyVoice. FSU-Mid gives the highest mean recovery ($100\%$) and the highest value in seven of the eight cells, with three network evaluations. FSU reaches $76\%$ on average, the same as the guidance interval of~\cite{kynk}, which coincides with FSU at two steps because its high-noise cutoff falls inside the first step. The only rule above FSU-Mid in any cell is the corrected TV-CFG on IndexTTS-2 in English ($82\%$ against $77\%$). On the uniform two-step grid of IndexTTS-2 its schedule sets the first-step weight to zero and raises the second to $1.4$, so it acts as FSU with a stronger second step, whereas on the cosine grid of the other decoders its first weight is $-0.15$ and it stays below FSU. This rule therefore also removes first-step guidance. Appendix~\ref{app:C} gives how each rule was ported and the SIM behind every cell.

\begin{figure}[t]
\centering
\includegraphics[width=\columnwidth]{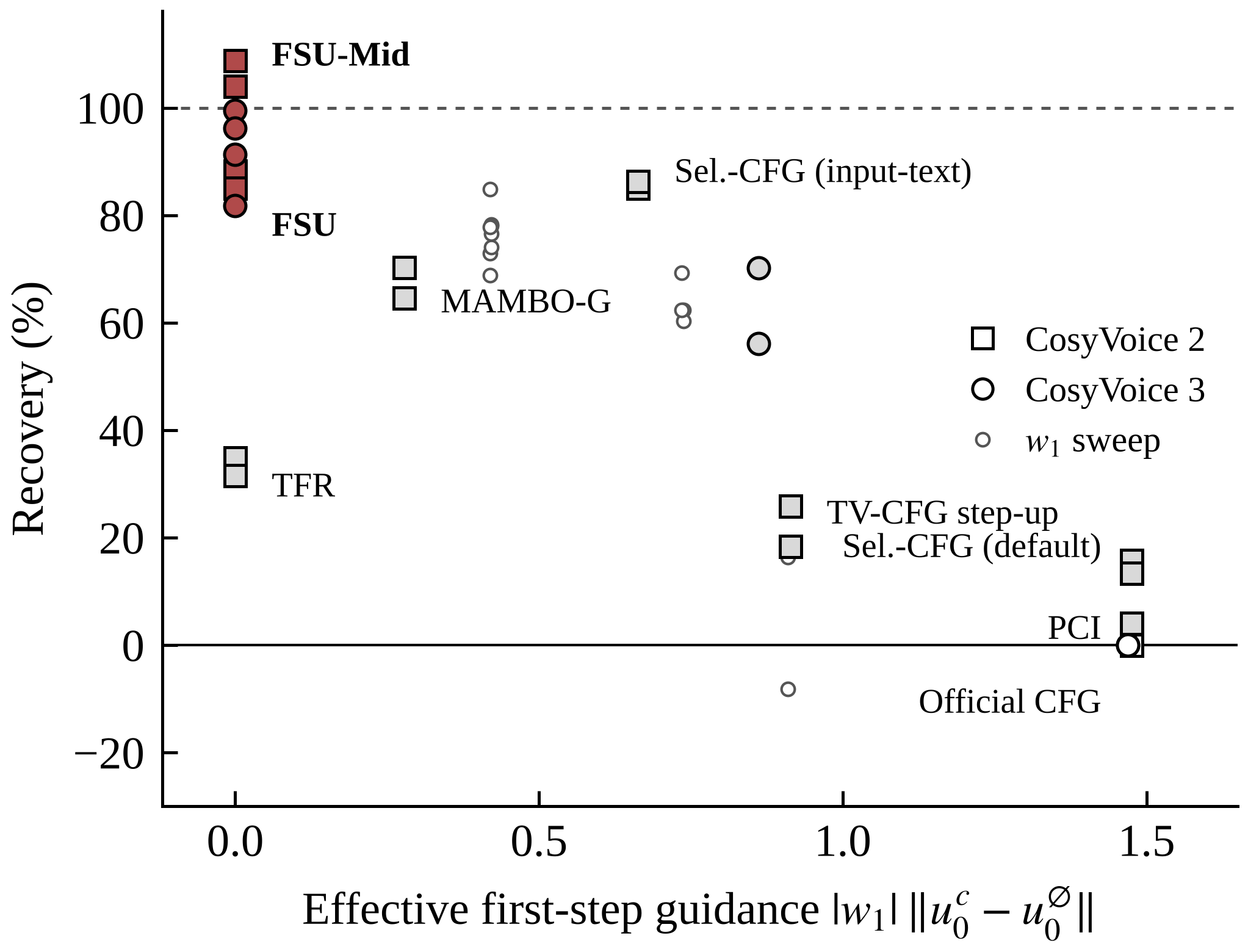}
\caption{Recovery against the guidance left on the first step, $|w_1|\,\|u^c_0-u^\emptyset_0\|$, on CosyVoice~2 and~3 at two steps. Red markers are FSU and FSU-Mid, gray markers the published rules and small circles the $w_1$ sweep.}
\label{fig:rulescatter}
\end{figure}

The best rule that keeps first-step guidance, Selective-CFG (input-text), reaches $64\%$ on average by shrinking $\|\gc\|$ through its speaker-null branch, and the rules that leave the first-step term unchanged recover at most $16\%$ in any cell. Two released implementations collapse on every decoder, and JRR-CFG falls below the official CFG while doubling the evaluations. On CosyVoice, recovery follows how much first-step guidance each rule leaves (Fig.~\ref{fig:rulescatter}, Spearman $-0.77$ on CosyVoice~2 and $-0.88$ on CosyVoice~3). The published rules and the $w_1$ sweep follow a single decreasing trend, and FSU and FSU-Mid lie at its upper end, where the first-step term is zero. The exception is terminal-fitted repair, which also leaves the first step unguided at two steps but modifies the last step and recovers only $32$--$35\%$.

\begin{table*}[t]
\caption{Full evaluation sets at two steps (Seed-TTS-eval test-en, test-zh and hard, LibriSpeech-PC). Recovery is (method $-$ official)/(teacher $-$ official), in parentheses for hard and LSPC. Bold marks the best few-step SIM per decoder with its official branch. All SIM differences to the official CFG are significant ($p\le10^{-3}$, Holm) except MAMBO-G on GLM-TTS. $^{\S}$Gap below seed noise, so the gain is given. $^\ast$CER worse than the official CFG ($p=0.01$).}
\label{tab:main}
\centering\footnotesize
\setlength{\tabcolsep}{4pt}\begin{tabular}{lcccccc}
\toprule
Method (NFE) & SIM en / zh & Recovery en / zh & SIM hard & SIM LSPC & WER / CER & UTMOS en / zh \\
\midrule
\multicolumn{7}{l}{\textit{CosyVoice 2}} \\
Official CFG (4) & 68.86 / 76.63 & 0\% / 0\% & 73.05 & 68.62 & 2.77 / 1.23 & 3.79 / 3.10 \\
TV-CFG (corrected)~\cite{stagewise} (4) & 72.93 / 79.47 & 70\% / 66\% & 77.03 (83\%) & 71.82 (62\%) & 2.85 / 1.28 & 4.05 / 3.22 \\
Sel.-CFG (input-text)~\cite{selcfg} (4) & 74.15 / 80.49 & 91\% / 90\% & 77.52 (94\%) & 74.06 (105\%) & 2.77 / 1.29 & 3.97 / 3.21 \\
MAMBO-G~\cite{mambog} (4) & 72.96 / 79.51 & 70\% / 67\% & 76.47 (72\%) & 72.31 (71\%) & 2.78 / 1.23 & 4.21 / 3.57 \\
FSU \ours\ (3) & 73.95 / 80.37 & 87\% / 87\% & 77.39 (91\%) & 73.23 (89\%) & 2.94 / 1.21 & 4.17 / 3.44 \\
FSU-Mid \ours\ (3) & \textbf{74.87} / \textbf{81.27} & 103\% / 108\% & \textbf{78.87} (122\%) & \textbf{74.49} (113\%) & 2.91 / 1.26 & 3.98 / 3.11 \\
Teacher (20) & 74.70 / 80.93 & 100\% / 100\% & 77.82 & 73.80 & 2.89 / 1.26 & 4.17 / 3.48 \\
\midrule
\multicolumn{7}{l}{\textit{CosyVoice 3}} \\
Official CFG (4) & 70.15 / 77.75 & 0\% / 0\% & 75.43 & 71.17 & 2.38 / 1.23 & 3.72 / 3.24 \\
TV-CFG (corrected)~\cite{stagewise} (4) & 75.02 / 81.96 & 72\% / 73\% & 80.14 (82\%) & 75.44 (68\%) & 2.45 / 1.10 & 3.87 / 3.12 \\
Sel.-CFG (input-text)~\cite{selcfg} (4) & 74.77 / 81.19 & 68\% / 59\% & 78.71 (57\%) & 75.17 (64\%) & 2.35 / 1.12 & 3.77 / 3.09 \\
MAMBO-G~\cite{mambog} (4) & 74.99 / 81.64 & 71\% / 67\% & 79.28 (67\%) & 75.31 (66\%) & 2.41 / 1.09 & 4.05 / 3.45 \\
FSU \ours\ (3) & 76.21 / 82.58 & 89\% / 84\% & 80.29 (85\%) & 76.61 (87\%) & 2.39 / 1.11 & 3.99 / 3.30 \\
FSU-Mid \ours\ (3) & \textbf{76.76} / \textbf{83.36} & 97\% / 97\% & \textbf{81.63} (108\%) & \textbf{77.69} (104\%) & 2.52 / 1.09 & 3.72 / 2.94 \\
Teacher (20) & 76.96 / 83.53 & 100\% / 100\% & 81.17 & 77.42 & 2.34 / 1.13 & 3.97 / 3.32 \\
\midrule
\multicolumn{7}{l}{\textit{IndexTTS-2}} \\
Official CFG (4) & 70.76 / 73.46 & 0\% / 0\% & 71.00 & 68.07 & 2.40 / 0.98 & 3.45 / 2.72 \\
TV-CFG (corrected)~\cite{stagewise} (4) & \textbf{75.98} / 79.11 & 81\% / 73\% & 76.77 (69\%) & 74.69 (80\%) & 2.40 / 0.94 & 3.44 / 2.43 \\
Sel.-CFG (input-text)~\cite{selcfg} (4) & 74.22 / 77.47 & 54\% / 52\% & 75.23 (50\%) & 67.08 ($-$12\%) & 2.48 / 1.00 & 3.34 / 2.45 \\
MAMBO-G~\cite{mambog} (4) & 72.93 / 76.57 & 34\% / 40\% & 74.43 (41\%) & 70.99 (35\%) & 2.42 / 0.99 & 3.58 / 2.70 \\
FSU \ours\ (3) & 74.31 / 77.91 & 55\% / 57\% & 75.47 (53\%) & 72.59 (54\%) & 2.44 / 0.97 & 3.43 / 2.47 \\
FSU-Mid \ours\ (3) & 75.74 / \textbf{79.65} & 77\% / 80\% & \textbf{77.78} (81\%) & \textbf{75.17} (85\%) & 2.48 / 0.97 & 3.41 / 2.45 \\
Teacher (50) & 77.19 / 81.23 & 100\% / 100\% & 79.41 & 76.38 & 2.54 / 1.01 & 3.65 / 3.00 \\
\midrule
\multicolumn{7}{l}{\textit{GLM-TTS}} \\
Official CFG (4) & 74.65 / 80.13 & $+0.00^{\S}$ / 0\% & 77.78 & 74.16 & 2.55 / 0.96 & 3.84 / 3.07 \\
TV-CFG (corrected)~\cite{stagewise} (4) & 75.60 / 80.90 & $+0.95^{\S}$ / 81\% & 78.75 (64\%) & 75.10 (164\%) & 2.50 / 0.98 & 3.90 / 3.13 \\
MAMBO-G~\cite{mambog} (4) & 74.72 / 80.14 & $+0.06^{\S}$ / 1\% & 77.85 (5\%) & 74.05 ($-$19\%) & 2.47 / 0.96 & 4.02 / 3.30 \\
FSU \ours\ (3) & 75.47 / 80.79 & $+0.82^{\S}$ / 70\% & 78.49 (47\%) & 74.88 (125\%) & 2.50 / 0.97 & 3.99 / 3.26 \\
FSU-Mid \ours\ (3) & \textbf{76.32} / \textbf{81.96} & $+1.67^{\S}$ / 195\% & \textbf{80.44} (175\%) & \textbf{75.87} (297\%) & 2.66 / 1.06$^\ast$ & 3.80 / 2.94 \\
All-null CFG (4) & 72.21 / 78.34 & 0\% / 0\% & 75.87 & 72.09 & 2.47 / 0.97 & 3.93 / 3.25 \\
All-null + FSU \ours\ (3) & 74.92 / 80.51 & 95\% / 79\% & 78.28 (70\%) & 74.25 (82\%) & 2.37 / 1.01 & 4.10 / 3.34 \\
All-null + FSU-Mid \ours\ (3) & 76.35 / 82.03 & 145\% / 135\% & 80.61 (138\%) & 75.79 (140\%) & 2.54 / 1.04 & 3.85 / 2.93 \\
Teacher (20) & 75.06 / 81.07 & $+0.41^{\S}$ / 100\% & 79.30 & 74.73 & 2.62 / 1.11 & 3.72 / 2.97 \\
\midrule
\bottomrule
\end{tabular}
\end{table*}

\subsection{Full Evaluation Sets}
Table~\ref{tab:rules} compares the methods with the published rules on the 300-utterance analysis subset and one noise seed. Table~\ref{tab:main} reports the official CFG, FSU and FSU-Mid on the complete Seed-TTS-eval test sets (test-en, test-zh and hard) and on LibriSpeech-PC for all four decoders at two steps, with WER and UTMOS, which is the standard reporting setting of zero-shot TTS. FSU recovers $84$--$89\%$ of the loss on CosyVoice~2 and~3 with no significant WER or CER increase and with UTMOS at the teacher's level. On IndexTTS-2, FSU-Mid recovers $77/80\%$, improving on FSU ($55/57\%$) on this uniform grid. On GLM-TTS, whose released speaker-null branch already keeps the loss small, both methods still raise SIM. FSU-Mid matches or exceeds the ten-step teacher on CosyVoice~2 and GLM-TTS on every set, with three network evaluations against twenty. The gains hold on the hard set ($91\%$ and $85\%$ recovery on CosyVoice~2 and~3) and on LibriSpeech-PC ($89\%$ and $87\%$ on CosyVoice~2 and~3). On LibriSpeech-PC, FSU leaves WER unchanged on CosyVoice~2, IndexTTS-2 and GLM-TTS ($p\ge0.06$), and on CosyVoice~3 its WER stays at the teacher's level.

The three best published rules of Table~\ref{tab:rules}, run on the same full sets with the same hyperparameters, keep their subset recovery on CosyVoice and IndexTTS-2 to within $6$ points. Selective-CFG (input-text) is omitted for GLM-TTS, where it equals the official speaker-null branch. FSU-Mid has the highest SIM in seven of the eight test-en and test-zh cells and in every hard and LibriSpeech-PC cell, and the exception is again the corrected TV-CFG on IndexTTS-2 in English ($+0.24$). Selective-CFG (input-text) is the strongest rule against FSU. It is level with FSU on CosyVoice~2 ($+0.20/+0.12$, below seed noise) but $1.4$ below it on CosyVoice~3, and on LibriSpeech-PC with IndexTTS-2 it falls below the official CFG. With the all-null branch on GLM-TTS, FSU-Mid exceeds the speaker-null teacher on all three Seed-TTS-eval sets and on LibriSpeech-PC.

\subsection{Robustness and Deployment}
Table~\ref{tab:robust} tests both samplers in sixteen conditions outside the analysis set. On CosyVoice, FSU recovers most of the loss across languages ($65$--$100\%$), with prompts from one to ten seconds and with noisy prompts at $20$ and $10$\,dB SNR ($86$--$113\%$), and on long inputs ($68$--$82\%$), where the official streaming mode keeps most of the gain of offline decoding. FSU-Mid reaches or exceeds the ten-step teacher on every long-input and prompt condition ($98$--$117\%$) and recovers $87$--$95\%$ on German, where FSU alone recovers $65$--$68\%$. On GLM-TTS German, FSU recovers $59\%$ and FSU-Mid $88\%$ with the all-null branch, and FSU-Mid exceeds FSU with the released branch, whose loss is already small. No condition shows a significant WER increase for either sampler.

The three best published rules were run in the same sixteen conditions. FSU-Mid lies above FSU in fourteen of them, the other two being within seed noise, and has the highest few-step SIM in twelve. MAMBO-G and the corrected TV-CFG stay below FSU in every condition but one and below FSU-Mid in all of them. Selective-CFG (input-text) is level with FSU in most conditions and ahead of it in four multilingual ones, where FSU-Mid closes most of the gap and exceeds it on CosyVoice~3. These gains come at the same wall-clock time as the official CFG (Table~\ref{tab:latency}).

%=====================================================================
\section{Applicability and Evaluation}
\label{sec:boundary}
\label{sec:boundaries}
The analysis predicts where FSU should fail and why FSU-Mid is needed. This section examines both predictions and the aspects of the output that SIM alone does not capture.

\subsection{Text-Conditioned Decoders}
As the analysis predicts, text-conditioned decoders behave differently. They have $\kappa_c>1$, so the conditional mean over all alignments is far from any single utterance and first-step guidance fixes the alignment. Removing it raises WER on F5-TTS from $2.33$ to $6.37$ at eight steps and on ZipVoice from $1.54$ to $2.45$ at four, with mixed SIM changes. For these decoders the official first-step guidance is best kept. Their official grids make the guided first step very short. With a uniform grid the F5-TTS baseline collapses at eight steps (WER $80.6\%$) and FSU repairs most of it (WER $18.9\%$), so a long guided first step from pure noise is harmful there as well.

\subsection{Truncation-Dominated Decoders}
The uniform grid of IndexTTS-2 makes the first step half of the trajectory, so this step carries both the first-step shift and part of the useful guidance. FSU alone recovers more than half of the loss, with a drop in predicted MOS (UTMOS and NISQA) in Chinese. FSU-Mid restores a guided evaluation at $t=0.5$, recovers $77$--$85\%$ and reduces this quality cost (Appendix~\ref{app:C}), which makes it the recommended sampler for such decoders.

\subsection{Evaluation Beyond Speaker Similarity}
\label{sec:eval}
A SIM close to the teacher's does not show that the teacher's distribution is reproduced. On CosyVoice~2 at two steps, Selective-CFG+FSU comes within seed noise of the teacher's SIM, yet its displacement $\Delta d_{\rm cond}$ stays at $0.018$ against $0.001$ for FSU and its embedding, F0 and speaking-rate variance ratios ($0.49/0.41/0.55$) stay at the level of Selective-CFG alone, and FSU-Mid exceeds the teacher's SIM with a slightly wider cloud. Among these samplers only FSU returns $\Delta d_{\rm cond}$ to the teacher's level while recovering most of the SIM loss, and the geometry of Section~\ref{sec:geometry} and the right panel of Fig.~\ref{fig:controls} are therefore reported alongside SIM. On quality, FSU is within seed noise of the teacher on UTMOS, UTMOSv2 and DNSMOS for the CosyVoice decoders and close to it on NISQA-TTS (within $0.21$ MOS, Appendix~\ref{app:C}).

%=====================================================================
%=====================================================================
\section{Conclusion}
Token-conditioned flow-matching TTS decoders lose speaker similarity at two or three steps mainly because of classifier-free guidance on the step that starts at pure noise. The resulting shift is nearly the same for every noise sample, persists through the remaining steps and scales with the first-step length. This offers an explanation for the speaker-similarity loss reported by few-step systems and for why guidance schedules designed for many steps do not remove it. In practice, few-step token-conditioned decoders can skip guidance on the first step, using FSU when the truncation error is small and FSU-Mid otherwise, with no training and one network evaluation fewer than the official two-step sampler. The conditioning branch and the ratio $\kappa_c$ indicate where this rule applies, and text-conditioned decoders, whose first step still sets the alignment, should keep the official guidance.

The experiments run public teachers with few Euler steps, and extending the analysis to distilled token-conditioned students~\cite{progdist,consistency} is left for when such checkpoints become public. Persistence of the first-step shift is established by measurement (Appendix~\ref{app:B}), and quality is assessed with five objective predictors rather than a listening test. The analysis itself depends only on the first Euler step starting at pure noise, so the same displacement should arise in any guided flow or diffusion sampler whose first step is long, and testing it on image and video generators is a natural next step.

%=====================================================================
\setcounter{topnumber}{4}\setcounter{bottomnumber}{4}\setcounter{totalnumber}{8}\renewcommand{\topfraction}{0.95}\renewcommand{\bottomfraction}{0.95}\renewcommand{\textfraction}{0.05}\renewcommand{\floatpagefraction}{0.8}
%=====================================================================
\appendices
%=====================================================================
\section{Experimental Details}
\label{app:A}

\textit{Models and data.} CosyVoice~2-0.5B uses a conformer-UNet decoder, IndexTTS-2 a DiT on a uniform grid, and GLM-TTS its own 1.5B LM with a flow DiT~\cite{dit} and the HiFT vocoder and CAM++ encoder~\cite{campp} of CosyVoice~2. The reproduced Fun-CosyVoice3-0.5B is $2.2$ SIM below the published value on test-en, so every comparison involving it is paired within the same pipeline. Robustness uses the MiniMax multilingual sets for ko/ja/de (100 each), LibriSpeech prompts of 1, 3 and 10\,s with MUSAN noise~\cite{musan}, and three-sentence concatenations for streaming. The analysis subset (300/300/100) agrees with the full sets to within $2.4$ percentage points of recovery ($4$ for GLM-TTS zh), and three noise seeds on the main cells (SIM s.d.\ $\le0.17$) change no ranking.

\textit{Statistics and seed noise.} Comparisons are paired with identical LM tokens, length and noise seed, the seed being derived from the utterance identifier and the seed index. Intervals are 95\% bootstrap intervals (10k resamples), tests are Wilcoxon signed-rank tests, and stars denote $p<0.05$ after Holm correction within each table. SIM removes the prompt region before embedding, and WavLM-SV~\cite{wavlm} ($\varepsilon=1.06$) and ECAPA~\cite{ecapa} ($0.71$) agree in sign with ERes2Net ($0.42$) in every comparison. The seed noise of the quality predictors is $0.06$--$0.16$ (UTMOS), $0.16$--$0.20$ (UTMOSv2), $0.05$--$0.10$ (DNSMOS), $0.14$--$0.26$ (NISQA) and $0.19$--$0.31$ (NISQA-TTS). A recovery is reported only when the teacher--official gap is at least $\varepsilon$, and otherwise the gain is reported. The teacher WER differs from published values by at most $0.3$, so all WER claims are paired, and because utterance-level WER differences are heavy-tailed the bootstrap interval of the mean and the Wilcoxon test are both reported (for FSU on CosyVoice~2 en, $2.77\to2.94$, interval $[-0.05,0.44]$, $p=0.60$).

\textit{Deployment.} Table~\ref{tab:latency} gives the latency of the official CosyVoice~2 streaming pipeline. Each chunk is solved once from noise, so the first-step effect does not accumulate across chunks, the recovery does not depend on the chunk size and no artifacts appear at chunk boundaries. FSU and FSU-Mid run within $2\%$ of the wall-clock time of the official CFG.

\begin{table*}[t]
\caption{Official CosyVoice~2 streaming pipeline (RTX~4080\,S, 50 requests). Throughput is in audio seconds per second at concurrency $c$. FSU-Mid throughput is scaled from FSU by their measured ratio ($1.00\pm0.01$).}
\label{tab:latency}
\centering\small
\begin{tabular}{lcccccc}
\toprule
Config & NFE & First packet (ms) & Decoder share & Thr.\ $c{=}1$ & $c{=}4$ & $c{=}16$ \\
\midrule
teacher, 10 steps & 20 & $1694\pm126$ & 37.7\% & 1.29 & 1.64 & 1.35 \\
baseline, 2 steps & 4 & $1244\pm134$ & 19.4\% & 1.57 & 2.17 & 1.95 \\
FSU \ours, 2 steps & 3 & $1249\pm133$ & 19.4\% & 1.57 & 2.13 & 1.94 \\
FSU-Mid \ours & 3 & $1237\pm139$ & 19.4\% & 1.59 & 2.12 & 1.93 \\
\bottomrule
\end{tabular}
\end{table*}

%=====================================================================
\section{Verification Details}
\label{app:B}
This appendix gives the measurements behind Sections~\ref{sec:remedy} and~\ref{sec:verify}.

\textit{Closed form (Section~\ref{sec:closedform}).} The network's guidance difference at $t=0$ is compared with $\hat\gc$, the mean of eight teacher samples minus the mean of eight unconditional samples at $N=10$. Its direction has cosine $0.96$--$0.98$ with $\hat\gc$, its norm is within $7$--$10\%$, and it varies little across seeds (mean cosine $0.986$--$0.998$). On 100 CosyVoice~2 utterances with eight seeds, the unguided one-step output lies $0.31$ mel RMS from the teacher mean, inside the teacher's own spread of $0.41$, and depends on the seed one tenth as much as the teacher. At the official $w=0.7$ it lies $1.36$ away, and at $w=1.5$ it lies $3.03$ away, beyond the unconditional mean at $2.27$.

\textit{Center distance (Section~\ref{sec:geometry}).} For each utterance, $\mu_f$ and $\mu_t$ are the means of eight few-step and eight teacher embeddings and $\sigma_t^2$ the mean squared distance of the teacher embeddings to $\mu_t$. The center distance $\delta^2=\max\{0,\|\mu_f-\mu_t\|^2/\sigma_t^2-(\sigma_f^2+\sigma_t^2)/(8\sigma_t^2)\}$ removes the expected contribution of the two center estimates, and two random halves of the teacher's seeds give the floor $\delta=0.35$.

\textit{Geometry of each sampler.} At two steps the official CFG has variance ratio $0.60/0.59$, $\Delta d_{\rm cond}=+0.043/+0.026$ and $\delta=1.28/1.34$ on CosyVoice~2 (en/zh), with similar values on CosyVoice~3 and $\delta$ up to $1.77$ on IndexTTS-2. FSU returns $\Delta d_{\rm cond}$ to $0.001$--$0.002$ on CosyVoice with the variance ratio unchanged and $\delta$ at $0.74$--$0.83$, and FSU-Mid comes closest to the teacher center ($\delta=0.56$--$0.89$ on all four decoders) with a variance ratio of $1.0$--$1.2$. The residual after FSU does not lie along the teacher's own guidance direction. With $a$ the vector from the unguided to the guided ten-step center, the residual has $1.5$--$11\%$ of its energy along $a$, against $21$--$58\%$ for the unguided two-step sampler.

\begin{figure}[t]
\centering
\includegraphics[width=\columnwidth]{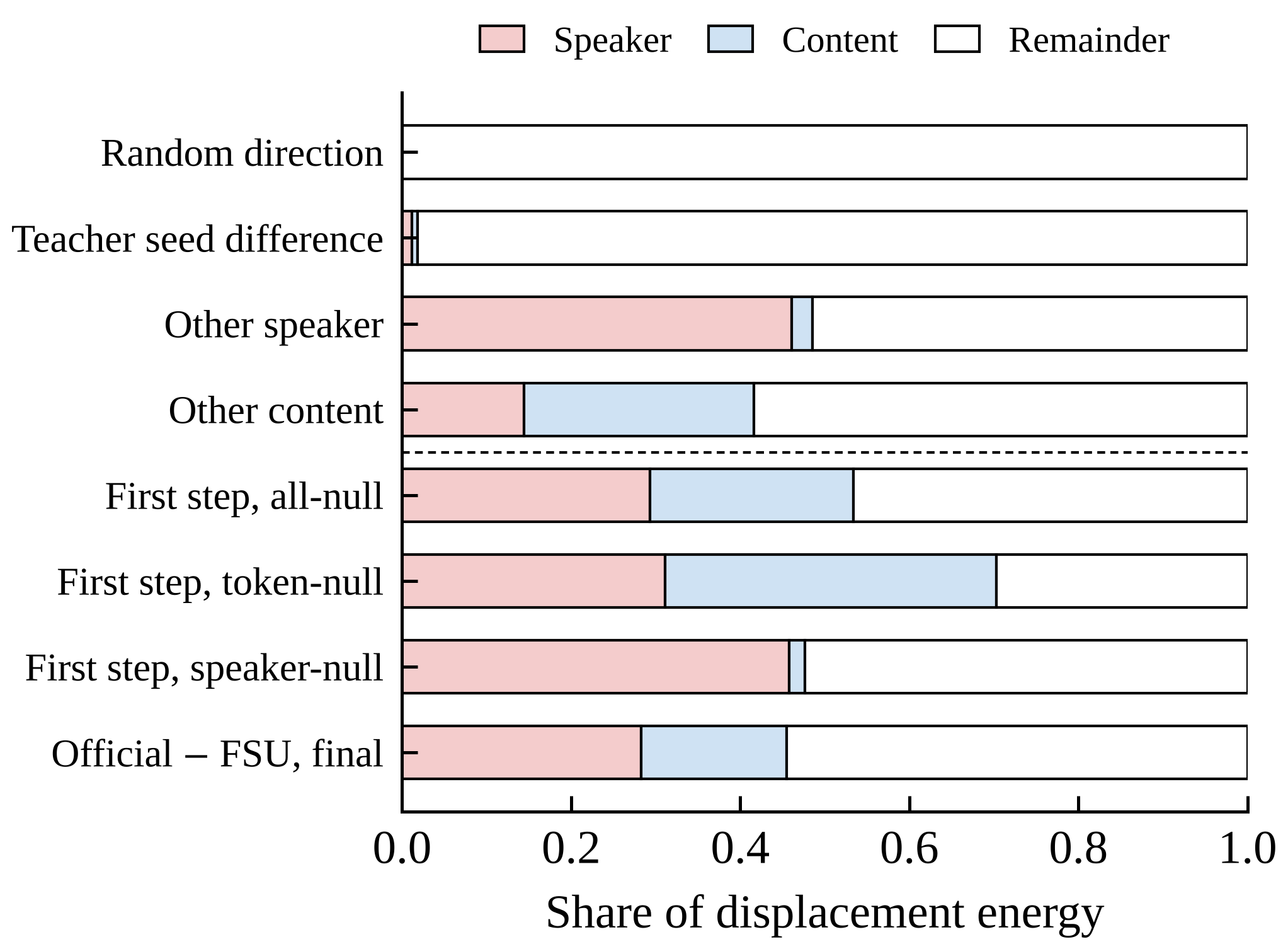}
\caption{Energy share of each displacement in speaker, content and remaining directions (CosyVoice~2, 100 utterances). Rows above the dashed line are controls.}
\label{fig:direction}
\end{figure}

\textit{Direction.} Fig.~\ref{fig:direction} decomposes each displacement in a per-utterance basis built from outputs for the same tokens under eight prompts. The first-step displacement puts $29\%$ of its energy in speaker directions, 24 times the share of the teacher's seed-to-seed difference and about $60\%$ of a genuine change of speaker.

\textit{Step length (Section~\ref{sec:law}).} The Spearman correlation of loss and $E_1$ is $0.88$, $0.97$, $0.98$ and $0.75$ for CosyVoice~2, CosyVoice~3, IndexTTS-2 and GLM-TTS. On CosyVoice~2 the FSU gain grows with the first-step length ($-0.1$, $+3.1$, $+5.2$, $+7.4$ and $+7.8$ for $t_1=0.1,\dots,0.7$). In $8/8$ model and language pairs, the list of cross-lingual prompts (prompt and target in different languages) with the larger $\|\gc\|$ had the larger loss and the larger gain ($p<10^{-6}$).

\textit{Attribution and threshold (Section~\ref{sec:controls}).} On CosyVoice~2, removing guidance from the first step recovers $88/85\%$ of the loss (en/zh), from both steps $35/32\%$, and lowering both weights to $0.3$ recovers $54/48\%$. The guidance magnitude $\gamma(t)$ falls to $0.31$--$0.36$ of its value at noise by $t=0.1$ and to $0.08$--$0.15$ by $t=0.5$ in all four decoders. The threshold $t^*=0.1$ is not tuned to the results. A sweep over $t^*$ puts it within $0.1$ SIM of the best value in all twelve cells of three decoders and four step counts. Consistently, FSU gains nothing on the grid $[0,0.1,1]$, where the guided second step starts at $0.31\gamma(0)$ and lasts $0.9$, and at four steps leaving steps~1 and~2 unguided, both starting below $t^*$, gains $+0.24/+0.67$ where step~1 alone gains nothing.

\textit{Two components of the loss.} The unguided sampler measures the Euler truncation error alone. From two steps to the official count its SIM changes by $-0.7/+0.7$ on CosyVoice~2 and $+0.7/+1.8$ on GLM-TTS but by $+2.1/+3.3$ on IndexTTS-2, whose uniform grid takes a first step of length $0.5$. Of the two-step loss of $6.2/7.6$ on IndexTTS-2, FSU removes $3.5/4.3$, against $5.1/3.6$ of $5.8/4.2$ on CosyVoice~2, and this larger residual is what FSU-Mid targets.

\textit{Branch and $\kappa_c$ (Section~\ref{sec:branch}).} On one-step outputs of the official samplers (en 100 $\times$ 2 seeds), $\hat r_c$ is the distance of the $w=0$ output to the nearest of eight teacher samples and $\|\hat\gc\|$ the RMS of the $w=1$ output minus the $w=0$ output (Table~\ref{tab:branch}). Absolute distances are in different feature spaces, so only the ratio $\kappa_c=\hat r_c/\|\hat\gc\|$ is compared, and it changes by at most $10.5\%$ with the number of teacher samples. For \eqref{eq:tower}, $\|g^{\rm all}\|^2$ is $15\%$ below the sum of its two components on CosyVoice, and the order all-null $\approx$ token-null $\gg$ speaker-null holds in all decoders.

\begin{table*}[t]
\caption{First-step guidance magnitude by unconditional branch (mel RMS), averaging term $\hat r_c$, ratio $\kappa_c$ for the official branch (bold) and two-step loss.}
\label{tab:branch}
\centering\small
\begin{tabular}{lcccccc}
\toprule
Decoder & all-null & token-null & speaker-null & $\hat r_c$ & $\kappa_c$ & 2-step loss \\
\midrule
CosyVoice 2 & \textbf{2.11} & 1.98 & 0.95 & 0.44 & 0.21 & 5.8 \\
CosyVoice 3 & \textbf{2.10} & 1.94 & 1.23 & 0.47 & 0.23 & 6.7 \\
IndexTTS-2 & \textbf{1.93} & 1.94 & 1.24 & 0.54 & 0.29 & 6.2 \\
GLM-TTS & 1.99 & 2.26 & \textbf{1.23} & 0.55 & 0.49 (0.28$^\dagger$) & 0.8 (3.4$^\dagger$) \\
\midrule
F5-TTS (text) & \textbf{1.52} & 0.89 & 1.25 & 1.58 & 1.09 & WER $+4.4^\ddagger$ \\
ZipVoice (text) & 1.58 & \textbf{0.85} & 1.29 & 1.58 & 1.97 & WER $+0.7^\ddagger$ \\
\bottomrule
\end{tabular}
\par\vspace{2pt}{\footnotesize $^\dagger$with the all-null branch patched in. For the text-conditioned decoders, token-null is text-null (text erased, speech prompt kept) and speaker-null is audio-null (prompt erased, text kept). $^\ddagger$WER change when first-step guidance is removed, mean of en and zh (\S\ref{sec:boundaries}).}
\end{table*}

\textit{Persistence (Section~\ref{sec:persist}).} At two steps the outputs with and without first-step guidance are the last-step prediction at two states $E_1$ apart, so a bound on their distance needs $h_2\tilde L<1$ for the Lipschitz constant $\tilde L$ of the guided field. Finite differences on CosyVoice~2 teacher paths give $h_2\tilde L=2.26$, because the velocity of a straight-path flow contains a $1/(1-t)$ factor, so the bound is vacuous and persistence is measured instead.

%=====================================================================
\section{Additional Results}
\label{app:C}

\textit{Guidance rules (Section~\ref{sec:rules}).} All rules are written in the convention of \eqref{eq:cfg}. None of the official implementations runs on a TTS decoder as is, and Selective-CFG, TFR, PCI and JRR-CFG have no public code, so each equation was ported into the CosyVoice sampler and then to IndexTTS-2 and GLM-TTS with the CosyVoice hyperparameters. With default settings every patched sampler reproduces the unpatched output within run-to-run GPU nondeterminism. PCI is undefined on IndexTTS-2, whose prompt region is held at zero and enters only as a separate input.

The released TV-CFG scheduler sets the last step's $\Delta t$ to zero in its normalization and gives per-step scales of $5.80$ and $13.93$ at two steps, and the corrected normalization gives $0.85$ and $2.05$. As a check of the porting, PCI on ZipVoice-Distill reproduces the published gain at four and eight steps (WavLM SIM $+0.30$ to $+1.09$ against a published $+0.4$ to $+1.4$). The rules keep their subset recovery on the full sets to within $6$ points.

\textit{Objective quality.} On the full sets, FSU is within $0.04$ of the teacher on UTMOS and within $0.13$ on UTMOSv2 and DNSMOS for CosyVoice~2 and~3, with NISQA-TTS residuals of $0.10$--$0.21$, and the official two-step CFG is $0.26$--$0.41$ below the teacher on UTMOS and UTMOSv2. On GLM-TTS both samplers exceed the teacher on UTMOS ($+0.08$ to $+0.29$). On IndexTTS-2 FSU is $0.22$--$0.65$ below the teacher on UTMOS and up to $1.0$ on NISQA, and FSU-Mid reduces the NISQA-TTS residual to $0.34$--$0.35$.

\textit{Robustness (Section~\ref{sec:experiments}).} Table~\ref{tab:robust} gives the sixteen robustness conditions with both samplers and the three best published rules. FSU-Mid exceeds FSU significantly on both German sets, on all four long-input conditions and with the ten-second prompt ($+0.8$ to $+1.9$, Holm $p\le0.003$), and is within seed noise of it on the short and noisy prompts, where the teacher--official gap is itself $1.8$--$6.5$. No sampler raises WER significantly over the official CFG in any condition where the language is supported, and with short and noisy prompts $\Delta d_{\rm cond}\le0.009$ after FSU in every cell.

\begin{table*}[t]
\caption{Robustness at two steps (ERes2Net SIM, 100 utterances per condition, seed 0). TV-CFG is the corrected version and Sel.-CFG the input-text version. Rec.\ is the recovery of FSU, with the gain given where the official loss is below seed noise. WER is official / FSU. Bold marks the highest few-step SIM. $^\S$Selective-CFG (input-text) uses the speaker-null branch, which is the official branch of GLM-TTS, so it equals the official CFG of the speaker-null row. $^\ddagger$The decoder or its LM does not handle the language (WER or CER above $55\%$ for every sampler, the teacher included), so WER is not compared.}
\label{tab:robust}
\centering\footnotesize
\setlength{\tabcolsep}{3.5pt}
\begin{tabular}{lccccccccc}
\toprule
Condition & Official & TV-CFG & Sel.-CFG & MAMBO-G & FSU \ours & FSU-Mid \ours & Teacher & Rec. & WER \\
\midrule
\multicolumn{10}{l}{\textit{CosyVoice 2}} \\
ko & 74.59 & 78.67 & 81.41 & 79.60 & 80.08 & \textbf{81.60} & 80.30 & 96\% & 3.95 / 3.08 \\
ja & 79.81 & 83.58 & \textbf{85.60} & 84.26 & 84.81 & \textbf{85.60} & 84.81 & 100\% & 10.53 / 10.58 \\
de & 67.75 & 71.26 & \textbf{74.13} & 71.18 & 71.56 & 73.34 & 73.62 & 65\% & 63.4 / 81.6$^\ddagger$ \\
long en, offline & 72.54 & 77.14 & 77.70 & 76.91 & 77.90 & \textbf{79.45} & 79.09 & 82\% & 2.54 / 2.75 \\
long zh, offline & 78.83 & 81.45 & 82.04 & 81.42 & 82.12 & \textbf{83.87} & 83.15 & 76\% & 1.46 / 1.37 \\
long en, streaming & 73.94 & 77.60 & 78.10 & 77.57 & 78.52 & \textbf{80.21} & 80.04 & 75\% & 2.51 / 2.90 \\
long zh, streaming & 79.74 & 81.60 & 82.72 & 81.81 & 82.54 & \textbf{84.40} & 83.85 & 68\% & 1.50 / 1.31 \\
prompt 1 s & 52.98 & 54.29 & \textbf{55.08} & 54.59 & 54.90 & 54.72 & 54.73 & 110\% & 6.84 / 7.39 \\
prompt 3 s & 67.82 & 70.86 & 72.22 & 71.05 & 72.21 & \textbf{72.57} & 72.35 & 97\% & 5.20 / 4.04 \\
prompt 10 s & 74.06 & 77.80 & 79.96 & 78.54 & 79.66 & \textbf{80.42} & 80.58 & 86\% & 7.75 / 8.32 \\
3 s, 20 dB & 65.47 & 68.02 & 69.56 & 68.54 & \textbf{69.77} & 69.66 & 69.29 & 113\% & 3.60 / 4.21 \\
3 s, 10 dB & 61.71 & 63.33 & 64.84 & 64.26 & 65.13 & \textbf{65.52} & 65.45 & 91\% & 4.14 / 3.93 \\
\multicolumn{10}{l}{\textit{CosyVoice 3}} \\
ko & 72.90 & 80.15 & 79.96 & 79.64 & 80.78 & \textbf{81.51} & 81.43 & 92\% & 2.68 / 2.42 \\
ja & 80.32 & 84.13 & \textbf{85.98} & 84.72 & 85.04 & 85.92 & 85.99 & 83\% & 56.0 / 56.0$^\ddagger$ \\
de & 76.73 & 78.47 & 79.43 & 79.21 & 79.80 & \textbf{80.66} & 81.23 & 68\% & 0.96 / 1.26 \\
\multicolumn{10}{l}{\textit{GLM-TTS}} \\
de & 76.81 & 76.63 & 76.81$^\S$ & 76.23 & 76.54 & \textbf{77.22} & 77.32 & $-0.27$ & 126 / 100$^\ddagger$ \\
\multicolumn{10}{l}{\textit{GLM-TTS (all-null)}} \\
de & 72.38 & 74.58 & \textbf{76.81}$^\S$ & 75.11 & 75.28 & 76.73 & 77.32 & 59\% & 78.2 / 77.9$^\ddagger$ \\
\bottomrule
\end{tabular}
\end{table*}

\textit{One step.} At one step, where leaving the first step unguided removes guidance altogether, the unguided sampler still gains $4.0$--$6.6$ SIM on CosyVoice and $2.2$--$3.2$ on IndexTTS-2 over the guided one-step sampler but loses $2.6$--$3.0$ on GLM-TTS, so the official guidance is kept there.

\section*{Acknowledgment}
The authors used Claude Fable~5.1 (Anthropic) to check the grammar of this manuscript. The authors reviewed all content and take full responsibility for it. This research was partly supported by the InnoCORE program of the Ministry of Science and ICT(\#N10260008) grant funded by the Korea Government(MSIT) and Korea Institute for Advancement of Technology(KIAT) grant funded by the Korea Government(MOTIR) (RS-2026-25585869, HRD Program for Industrial Innovation).

\bibliographystyle{IEEEtran}
\bibliography{reference}

\end{document}